\documentclass[aps,pra,reprint,superscriptaddress]{revtex4-1}
\usepackage{amsmath}
\usepackage{bbm}
\usepackage{graphicx}
\usepackage{amsfonts}
\usepackage{color}
\usepackage{upgreek}
\makeatletter
\def\captionof#1#2{{\def\@captype{#1}#2}}
\makeatother
\usepackage{hyperref}
\hypersetup{
    colorlinks = true,
    linkcolor =blue,
	citecolor=blue, 
	urlcolor=blue 
}

\newcommand*{\defeq}{\mathrel{\vcenter{\baselineskip0.5ex \lineskiplimit0pt
                     \hbox{\scriptsize.}\hbox{\scriptsize.}}}%
                     =}

\begin{document}

\title{Periodically refreshed baths to simulate open quantum many-body dynamics
}

\author{Archak Purkayastha}
\email{archak.p@tcd.ie}
\affiliation{School of Physics, Trinity College Dublin, College Green, Dublin 2, Ireland}

\author{Giacomo Guarnieri}
\email{giacomo.guarnieri@fu-berlin.de}
\affiliation{School of Physics, Trinity College Dublin, College Green, Dublin 2, Ireland}
\affiliation{Dahlem Center for Complex Quantum Systems, Freie Universit at Berlin, 14195 Berlin, Germany}

\author{Steve Campbell}
\email{steve.campbell@ucd.ie}
\affiliation{School of Physics, University College Dublin, Belfield, Dublin 4, Ireland}
\affiliation{Centre for Quantum Engineering, Science, and Technology, University College Dublin, Belfield, Dublin 4, Ireland}

\author{Javier Prior}
\email{javier.prior@um.es}
\affiliation{Departamento de F\'isica, Universidad de Murcia, Murcia E-30071, Spain}
\affiliation{Instituto Carlos I de F\'isica Te\'orica y Computacional, Universidad de Granada, Granada 18071, Spain}

\author{John Goold}
\email{gooldj@tcd.ie}
\affiliation{School of Physics, Trinity College Dublin, College Green, Dublin 2, Ireland}

\date{\today}

\begin{abstract}
Obtaining dynamics of an interacting quantum many-body system connected to multiple baths initially at different, finite, temperatures and chemical potentials is a challenging problem. This is due to a combination of the prevalence of strong correlations in the system, the infinite nature of
the baths and the long time to reach steady state.
In this work we develop a general formalism that allows access to the full non-Markovian dynamics of such open quantum many-body systems up to the non-equilibrium steady state (NESS), provided its uniqueness.  Specifically, we show how finite-time evolution in presence of finite-sized baths, whose opportune size is determined by their original spectral density, can be recursively used to faithfully reconstruct the exact dynamics without requiring any small parameter. Such a reconstruction is possible even in parameter regimes which would otherwise be inaccessible by current state-of-the-art techniques. We specifically demonstrate this by obtaining the full numerically exact non-Markovian dynamics of interacting fermionic chains in two terminal set-ups with finite temperature and voltage biases, a problem which previously remained outstanding despite its relevance in a wide range of contexts, for example, quantum heat engines and refrigerators.

\end{abstract}

\maketitle

\section{Introduction}
Accurately obtaining dynamics and non-equilibrium steady states of a large complex many-body system in the presence of two or more baths at different finite temperatures and chemical potentials is a notoriously difficult problem to solve, despite its broad applicability in diverse fields such as quantum thermodynamics~\cite{goold2016role,benenti2017fundamental}, mesoscopic physics~\cite{datta1997electronic,akkermans2007mesoscopic}, quantum biology~\cite{lambert2013quantum}, quantum chemistry \cite{Molecular_junction_review}. In the absence of any small parameter and beyond quadratic Hamiltonians \cite{landauer1970electrical,buttiker1986four,benenti2017fundamental,Abhishek_Diptiman2006}, a solution for the open quantum system's dynamics is often intractable. This is due to a combination of many-body correlations in the system, the infinite degrees of freedom in the environments (baths), the non-Markovian nature of the dynamics \cite{deVega_2017_review} and the long time required to reach steady state.  Here we develop a general formalism which allows full reconstruction of such open-system dynamics by, instead, recursively using evolution up to a finite time, requiring only finite-size baths. 
When combined with state-of-the-art
numerical techniques \cite{thermofield_bosons,TEDOPA_bosons,TEDOPA_fermions,
Tamascelli_PRL,Makri_1995_I,Makri_1995_II,TEMPO,process_tensor,Schwarz_2018,
Boulat_2008,Zwolak_2020,TTM0,TTM}, it drastically simplifies the numerical simulation, allowing access to open quantum many-body dynamics in parameter regimes which have thus far remained intractable.

In particular, owing to limitations in practical implementation, all existing numerically-exact techniques for simulating non-Markovian dynamics \cite{deVega_2017_review,Makri_1995_I,Makri_1995_II,TEMPO,process_tensor,Schwarz_2018,
Boulat_2008,thermofield_bosons,TEDOPA_bosons,TEDOPA_fermions,
Tamascelli_PRL,Zwolak_2020,TTM0,TTM} have been hitherto limited to zero dimensional systems (e.g. impurity models, single three level systems, single qubits etc.). We demonstrate that our formalism allows to bypass these practical limitations and adopt these techniques for efficiently simulating long time non-Markovian dynamics of one-dimensional systems, provided there exists a unique non-equilibrium steady state (NESS). We show this by simulating the dynamics of interacting quantum many-body fermionic chains strongly coupled to two baths at different (finite) temperatures and chemical potentials, using one of these techniques \cite{TEDOPA_bosons,Tamascelli_PRL,TEDOPA_fermions}. Obtaining the numerically exact dynamics of interacting quantum many-body chains in such two-terminal set-ups has been an outstanding problem, despite its relevance in a wide range of contexts, such as quantum transport, localization, integrability breaking \cite{Marko, Znidaric_2011_XXZ,mendoza2015coexistence, vznidarivc2016diffusive,vznidarivc2017dephasing,PhysRevB.99.094435,schulz2020phenomenology,vznidarivc2018interaction,PhysRevB.100.085105,brenes2018high,PhysRevLett.125.180605}, quantum heat engines and refrigerators \cite{benenti2017fundamental}. Further, we discuss the relationship between our formalism and collisional (or repeated interaction) models \cite{RauPR, ScaraniPRL, ZimanPRA, ciccarello2017, Campbell_2021,BarraSciRep, GabrieleNJP2018, StrasbergPRX, Guarnieri2020PhysLettA,CiccarelloPRA, VacchiniPRL, StrunzPRA, CampbellPRA, CompositeCMs, CompositeCMs2,Cattaneo_2021}, highlighting how our results extend these notions, significantly advancing this highly active field of research. Finally, to demonstrate that our formalism can be combined with not one but any of the existing techniques for numerically exact non-Markovian dynamics \cite{deVega_2017_review,Makri_1995_I,Makri_1995_II,TEMPO,process_tensor,Schwarz_2018,
Boulat_2008,thermofield_bosons,TEDOPA_bosons,TEDOPA_fermions,
Tamascelli_PRL,Zwolak_2020,TTM0,TTM}, we also apply our formalism to a spin-boson model employing a completely different numerical technique \cite{TEMPO} compared to the one used for the many-body chains. 

The paper is arranged as follows. In Sec.~\ref{Sec:set-up}, we introduce the general set-up and assumptions. In Sec.~\ref{Sec:main_statement} we present and discuss our main statement. In Sec.~\ref{Sec:finite_baths}, we present how our main statement allows the use of finite size baths, without essentially any further approximation. In Sec.~\ref{Sec:collisional_models}, we discuss the connection to collisonal (or repeated interaction) models.  In Sec.~\ref{Sec:numerical_results}, we present our numerical demonstrations. In Sec.~\ref{Sec:conclusions}, we give our conclusions and outlook. This is followed by an Appendix, which contains the proof of our main statement (Appendix~\ref{app:derivation}), a discussion on what controls the memory time of baths (Appendix~\ref{app:estimating_memory}), exact analytical steady state results in case of quadratic (non-interacting) Hamiltonians (Appendix~\ref{app:NEGF}) which are used to benchmark results in Sec.~\ref{Sec:numerical_results}, explicit details of the numerical technique used to simulate the interacting fermionic chains (Appendix~\ref{app:TEBD_mixed_basis}).

\section{The set-up}\label{Sec:set-up}
We consider the general set-up of a quantum system connected to multiple baths.
The Hamiltonian generating the dynamics of the full set-up is denoted by 
\begin{align}
\hat{\mathcal{H}}=\hat{\mathcal{H}}_S + \hat{\mathcal{H}}_{SB} + \hat{\mathcal{H}}_B,
\end{align}
with 
$\hat{\mathcal{H}}_S$ being the Hamiltonian of the system, $\hat{\mathcal{H}}_B$ being the composite Hamiltonian of all the baths, and finally $\hat{\mathcal{H}}_{SB}$ being the interaction Hamiltonian between the system and all the baths. The initial state of the full set-up is taken to be of the product form $\hat{\rho}_{\rm tot}(t_0)=\hat{\rho}(t_0) \hat{\rho}_B$, where $\hat{\rho}_B$ is the composite initial state of all the baths, and $t_0$ is the initial time. The state of the system at a later time $t$ is given by,
\begin{align}
\label{def_Lambda0}
&\hat{\rho}(t)= \hat{\Lambda}(t-t_0)[\rho(t_0)] \nonumber \\
&\hat{\Lambda}(t-t_0)[\rho(t_0)] = {\rm Tr}_B\Big(e^{-i\hat{\mathcal{H}}(t-t_0)}\hat{\rho}(t_0) \hat{\rho}_B e^{i\hat{\mathcal{H}}(t-t_0)} \Big),   
\end{align}
where ${\rm Tr}_B(...)$ denotes trace over the bath degrees of freedom. The linear superoperator $\hat{\Lambda}(t-t_0) :\hat{\rho}(t_0) \to \hat{\rho}(t) $ is given by a completely-positive and trace preserving (CPTP).  Without loss of generality, we assume ${\rm Tr}_B(\hat{\mathcal{H}}_{SB} \hat{\rho}_B)=0$ \cite{Breuer_book}. The overarching goal of open quantum systems theory is to obtain the system's state $\hat{\rho}(t)$ at all times as accurately as possible \cite{Breuer_book,weiss2012quantum,Jauho_book,Kamenev_book}. This problem is intractable in complete generality, so we make some additional mild physical assumptions. Let us assume that the dynamics of the full set-up remains analytic at all times, and that the system Hilbert space is finite-dimensional. Note that, even if the system is a lattice of bosonic sites, but there is some effective cut-off on the number of bosons at each site (e.g. due to repulsive interactions or due to temperatures constraints in the problem) this amounts to considering an effective finite system Hilbert space dimension. We restrict to cases where the NESS is unique, i.e, the long time state of the system, $\hat{\rho}_{\rm NESS}$, is independent of the initial state, $\hat{\rho}(t_0)$. Many situations of interest involving quantum many-body systems fall under this class \cite{goold2016role,benenti2017fundamental,datta1997electronic,
akkermans2007mesoscopic,lambert2013quantum,Molecular_junction_review,deVega_2017_review}. On physical grounds, this requires that the system-size is finite (but can be large), while the baths are in the thermodynamic limit \cite{Purkayastha_2019}. An effective time to reach steady state $t_{\rm ss}$ can be defined, which satisfies  $|| \hat{\rho}_{\rm NESS} - \hat{\rho}(t) || < \epsilon,~\forall~t \geq t_{\rm ss}$, where $\epsilon$ is an arbitrary small number set by experimental or numerical precision, and $||\hat{O}||$ denotes operator norm of $\hat{O}$. The dynamics of the system will be non-Markovian in general. However, it can be shown that uniqueness of steady state necessitates an effective finite memory time $\tau_M$ for the dynamics of the system, consistent with physical intuition (Appendix~\ref{app:derivation}).

When $\hat{\mathcal{H}}_S$ describes an interacting quantum many-body system, even with above rather mild assumptions, it remains an extremely challenging problem. Further assumptions are most often made to enable a Markovian description of the system's dynamics \cite{michel2003fourier,prosen2008third,Marko, Znidaric_2011_XXZ,mendoza2015coexistence, vznidarivc2016diffusive,vznidarivc2017dephasing,PhysRevB.99.094435,schulz2020phenomenology,vznidarivc2018interaction,PhysRevB.100.085105,brenes2018high,PhysRevLett.125.180605,Xu_2019,Nathan_Rudner_2020}. These descriptions are limited to either weak system-bath couplings or to infinite temperatures \cite{Breuer_book,Gorini1978}. Non-Markovian descriptions beyond these regimes typically rely on diagrammatic perturbation theories requiring a small parameter \cite{Jauho_book,Kamenev_book,Stan2009,Wang2014}. 
On the other hand, a number of numerical techniques exist which allow numerically exact calculation of $\hat{\rho}(t)$, without requiring any small parameter, but are limited by the time up to which they can simulate \cite{Makri_1995_I,Makri_1995_II,TEMPO,process_tensor,thermofield_bosons,TEDOPA_bosons,
TEDOPA_fermions,Schwarz_2018,Boulat_2008,Tamascelli_PRL}. These techniques have therefore been limited to describe systems where $t_{\rm ss}$ is small or when only the short time dynamics is of interest. It is this drawback of this class of techniques that is removed by our formalism,  thereby allowing their use in cases previously deemed impossible. We call our formalism the \textit{Periodically Refreshed Baths} (PReB) approach.

\section{The main statement of ${\rm \bf PReB}$}\label{Sec:main_statement}

Given the set-up described in the previous section, our main result can be stated succinctly as follows.
\begin{align}
\label{PReB_statement0}
&\hat{\rho}_{n\tau+t_1} \defeq \underbrace{\hat{\Lambda}(\tau) [\ldots[\hat{\Lambda}(\tau)[}_{\text{n times}}\hat{\Lambda}(t_1-t_0)[\hat{\rho}(t_0)]]]\ldots],\nonumber \\
& \Big | \Big | \hat{\rho}(n\tau+t_1) - \hat{\rho}_{n\tau+t_1} \Big | \Big | = \epsilon(\tau), \textrm{ $\epsilon(\tau)$ decays with $\tau$}, 
\end{align}
where $t_0\leq t_1 < \tau$, and $\hat{\Lambda}(t)$ and $\hat{\rho}(t)$ are as defined before (Eq.(\ref{def_Lambda0})). The proof of this statement requires no further assumptions than already mentioned. The proof is given in Appendix~\ref{app:derivation}. The density matrix $\hat{\rho}_{n\tau+t_1}$, by definition, is obtained from the following physical process. Starting from a product state between the system and the baths, the system is evolved in presence of the baths up to time $t_1$. At this point, the baths are detached and refreshed to their original initial state, thus making the state of the full set-up again a product state of system and baths on the following step. Afterwards, this detaching and refreshing of the baths is done periodically in steps to time $\tau$. We dub this process PReB. Eq.(\ref{PReB_statement0}) says that, with increasing $\tau$, the state of the system obtained from the PReB process converges to the state obtained by a continuous time evolution up to $n\tau+t_1$, i.e, $\hat{\rho}(n\tau+t_1)$.  It can be shown that the $\tau$ required for convergence satisfies $t_{\rm ss} \geq \tau \gg \tau_M$. In many cases, the memory time $\tau_M$ is governed by the time for decay of correlations in the baths, irrespective of the system. On the other hand, if our set-up describes a quantum many-body system on a lattice connected to baths at only few sites, the time to reach steady state, $t_{\rm ss}$, will depend crucially on the internal dynamics of the system. If the system size is large but finite, typically $t_{\rm ss}$ will be orders of magnitude larger than $\tau_M$. In all such cases, on physical grounds, we expect to find convergence of $\hat{\rho}_{n\tau+t_1}$ with a $\tau$ satisfying 
\begin{align}
\label{tau_condition0}
t_{\rm ss} \gg \tau \gg \tau_M.
\end{align}
It is in these cases that Eq.(\ref{PReB_statement0}) becomes extremely useful because it implies that the long time evolution in the presence of the baths can be reconstructed by repeatedly using a simulation of much a shorter time evolution. Crucially, reconstruction up to any desired precision in time-steps is possible by simply re-running the process with different choices of the initial time step $t_1$. If, however, the $\tau$ required for convergence approaches $t_{\rm ss}$, which, for example, might happen for small systems at low temperatures with all sites strongly coupled to baths, Eq.(\ref{PReB_statement0}) will not offer any advantage over directly using existing numerical techniques \cite{Boulat_2008,Anders_2008,Eckel_2010,Cohen_2014,Dorda_2015,
Schwarz_2018,Chen_2019,Loten_2020,TEMPO,TTM0,TTM,process_tensor,Makri_1995_I,
Makri_1995_II}.

The state $\hat{\rho}_{n\tau+t_1}$ is obtained by a discrete time Markovian evolution in steps of $\tau$, starting from the state at time $t_1$. If the $\tau$ required for convergence is much smaller than all internal time scales of the system, the continuous limit of the Markovian evolution can be taken, and a quantum master equation in Lindblad form \cite{lindblad1976generators,Gorini1978}  can be derived. In these special cases, the dynamics of the system is effectively Markovian. Generically, however, the $\tau$ required for convergence is likely to be larger than at least some internal time scale of the system. So the continuous limit cannot be taken and it rules out the possibility to describe the system dynamics in terms of a Lindblad equation. In this sense, in generic cases, the results obtained from the converged PReB process represent non-Markovian dynamics of the system in the original set-up.

\section{Finite size baths via chain-mapping}\label{Sec:finite_baths}
\begin{figure}
\includegraphics[width=\columnwidth]{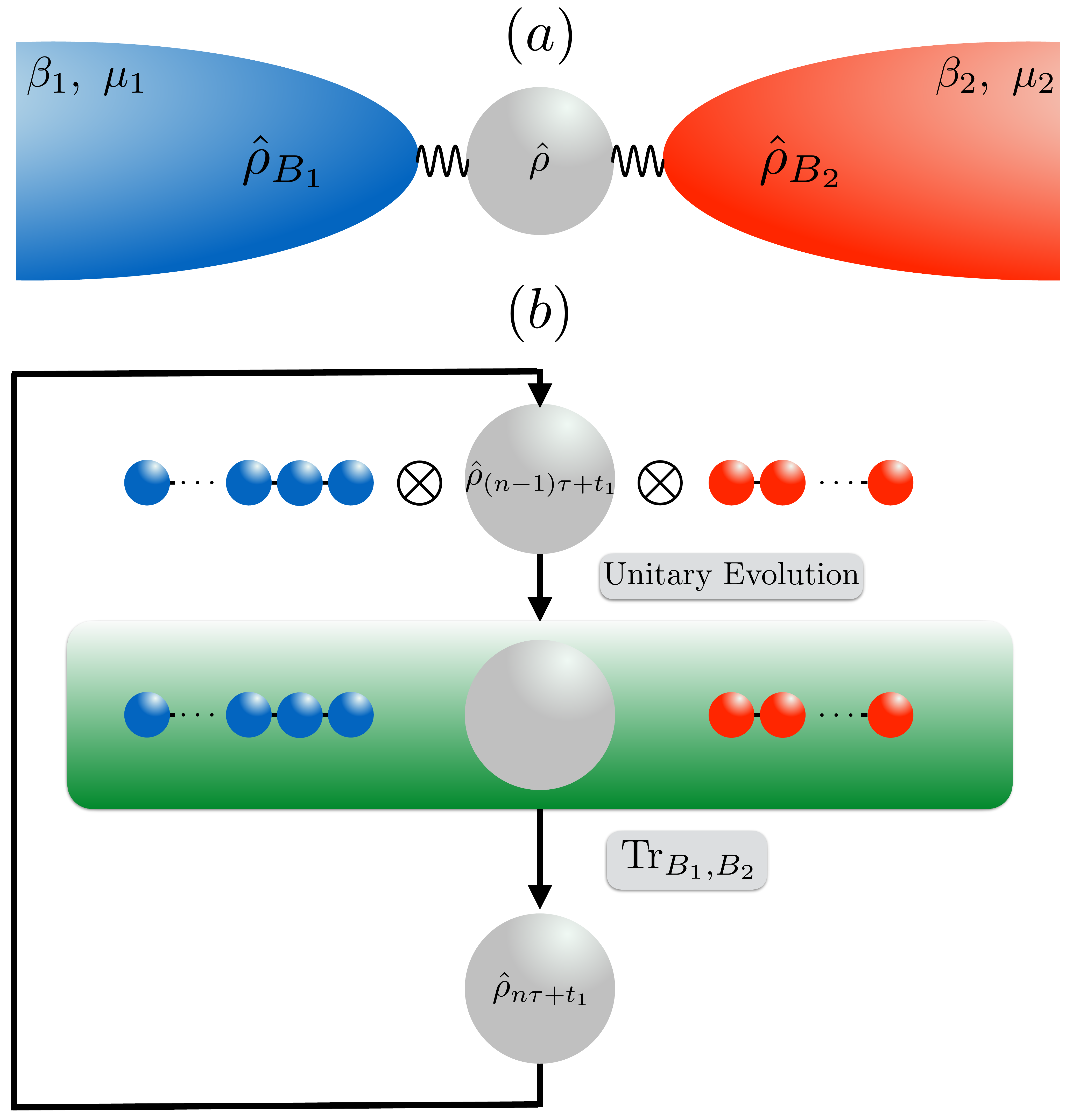} 
\caption{(a) The figure shows a schematic of a typical out-of-equilibrium set-up where the system, initially in an arbitrary state, is connected to two baths at different temperatures and chemical potentials. (b) The figure shows the $n$th step of Periodically Refreshed Baths (PReB) algorithm with two baths.  }
\label{fig:PReB_schematic} 
\end{figure}

The main statement of PReB is completely general, valid for arbitrary choices of system, bath and system-bath coupling Hamiltonians, as long as there is a unique NESS for the system. Let us now consider a more specific case, the canonical model for thermal baths, 
\begin{align}
&\hat{\mathcal{H}}_B~=~\sum_{\ell} \hat{\mathcal{H}}_B^{(\ell)},~ 
\hat{\mathcal{H}}_B^{(\ell)} =    \sum_{r=1}^{\infty} \Omega_{r\ell} \hat{B}_{r\ell}^\dagger \hat{B}_{r\ell}, \nonumber \\
& \hat{\mathcal{H}}_{SB}=\sum_\ell \hat{\mathcal{H}}_{SB}^{(\ell)},
~ \hat{\mathcal{H}}_{SB}^{(\ell)}=\sum_{r=1}^\infty(\kappa_{r\ell} \hat{S}^\dagger_\ell \hat{B}_{r\ell} + \kappa_{r\ell}^* \hat{B}_{r\ell}^\dagger\hat{S}_\ell ),
\end{align}
where $\hat{S}_\ell$ is some system operator coupling to the $\ell$th bath, $\hat{B}_{r\ell}$ is the fermionic or bosonic annihilation operator of the $r$th mode of the $\ell$th bath. The composite initial state of the baths is given by \begin{align}
\hat{\rho}_B=\prod_{\ell}\frac{e^{-\beta_\ell(\hat{\mathcal{H}}_B^{(\ell)}-\mu_\ell \hat{N}_B^{\ell})}}{Z_B^{(\ell)}},
\end{align}
where $\hat{N}_B^{\ell}$ is the total particle number operator of the $\ell$th bath, $Z_B^{(\ell)}$ is the corresponding partition function. For such baths, the influence of the baths on the dynamics of the system is entirely governed by the bath spectral functions, defined as 
\begin{align}
\mathfrak{J}_\ell(\omega) = 2\pi\sum_{r=1}^\infty |\kappa_{r\ell}|^2 \delta (\omega - \Omega_{r\ell}),
\end{align} 
and the Fermi or Bose distribution corresponding to the initial states of the baths, $\mathfrak{n}_\ell(\omega)=[{\rm exp}(\beta_\ell(\omega-\mu_\ell))\pm 1]^{-1}$. The effective memory time $\tau_M$ is given by the time for decay of the Fourier transforms of the functions $\mathfrak{J}_\ell(\omega)$ and $\mathfrak{J}_\ell(\omega)\mathfrak{n}_\ell(\omega)$, and thus is a property of the baths, independent of the system \cite{Ahana_2018,Nathan_Rudner_2020} (Appendix~\ref{app:estimating_memory}), as previously mentioned. Any bath spectral function $\mathfrak{J}_\ell(\omega)$ with finite upper and lower cut-offs in frequency can be exactly mapped onto a semi-infinite nearest neighbour non-interacting tight-binding chain with the first site coupled to the system \cite{rc_mapping_QTD_book,rc_mapping_bosons,rc_mapping_fermions,
rc_mapping_old,TEDOPA_bosons,TEDOPA_fermions,Chain_mapping_bosons_fermions}, 
\begin{align}
& \hat{\mathcal{H}}_B^{(\ell)}~=~\sum_{p=1}^{\infty} \left(\varepsilon_{p,\ell} \hat{b}_{p,\ell}^\dagger\hat{b}_{p,\ell} + g_{p,\ell}(\hat{b}_{p,\ell}^\dagger\hat{b}_{p+1,\ell}+\hat{b}_{p+1,\ell}^\dagger\hat{b}_{p,\ell})\right), \nonumber \\
& \hat{\mathcal{H}}_{SB}^{(\ell)}=\gamma_{\ell}(\hat{b}_{1,\ell}^\dagger \hat{S}_\ell  + \hat{S}^\dagger_\ell \hat{b}_{1,\ell}).
\end{align} 
The parameters $\gamma_{\ell}$,  are given by
\begin{align}
\gamma_{\ell}^2 = \frac{1}{2\pi} \int d\omega~\mathfrak{J}_\ell(\omega).
\end{align} 
The on-site potentials $\varepsilon_{p,\ell}$ and the hoppings $g_{p,\ell}$ are obtained from the following set of recursion relations
\begin{align}
& \mathfrak{J}_{p, \ell}(\omega)= \frac{4g_{p-1,\ell}^2 \mathfrak{J}_{p-1,\ell}(\omega) }{\left[\mathfrak{J}_{p-1,\ell}^H (\omega)\right]^2 +\left[\mathfrak{J}_{p-1,\ell}(\omega)\right]^2},  \nonumber \\
& g_{p, \ell}^2 = \frac{1}{2\pi} \int d\omega \mathfrak{J}_{p,\ell}(\omega), \\
& \varepsilon_{p, \ell} = \frac{1}{2\pi g_{p, \ell}^2} \int d\omega~\omega \mathfrak{J}_{p,\ell}(\omega) \nonumber,
\end{align}
with $\mathfrak{J}_{0,\ell}(\omega) =\mathfrak{J}_{\ell}(\omega) $ and $\mathfrak{J}_{p,\ell}^H (\omega)$ being the Hilbert transform of  $\mathfrak{J}_{p,\ell} (\omega)$,
\begin{align}
\mathfrak{J}_{p,\ell}^H(\omega)= \frac{1}{\pi}\mathcal{P}\int_{-\infty}^{\infty} d\omega^\prime \frac{\mathfrak{J}_{p,\ell}(\omega^\prime)}{\omega -\omega^\prime},
\end{align} 
where $\mathcal{P}$ denotes the principal value \cite{rc_mapping_QTD_book}.
The spectral function is now encoded in the on-site energies $\varepsilon_{p,\ell}$ and the hopping parameters $g_{p,\ell}$, and the strength of system-bath coupling $\gamma_{\ell}$. In particular,  $\varepsilon_{p,\ell}$, $g_{p,\ell}$, quickly tend to a constant with increase in the index $p$~\cite{rc_mapping_QTD_book}. We let the constants be $\varepsilon_{B_\ell}$, $g_{B_\ell}$. The value of $g_{B_\ell}$ is directly proportional to the bandwidth of the bath~\cite{rc_mapping_QTD_book}. After mapping the baths to tight-binding chains, due to Lieb-Robinson bounds, sites of the $\ell$th bath further than $\sim (t-t_0)g_{B_\ell}$ have a negligible effect on  dynamics of the system up to time $t$~\cite{Woods_2015,Woods_2016, Chain_mapping_bosons_fermions}. Therefore, to accurately simulate the process described by $\hat{\Lambda}(\tau)$, one needs the $\ell$th bath to be modelled by a chain of size $L_B\sim \tau g_{B_\ell}$. Thus, due to Eq.(\ref{PReB_statement0}) and Eq.(\ref{tau_condition0}),  reconstruction of the full dynamics by simulating the PReB process requires only finite-sized baths. Fig.~\ref{fig:PReB_schematic} demonstrates the PReB algorithm in presence of two baths after chain mapping, envisaged as a numerical method. Given any code to numerically exactly obtain the state of the system after time $\tau$ starting from an arbitrary initial state, the code can be used recursively to simulate the PReB process, as shown in Fig.~\ref{fig:PReB_schematic}(b).

\section{${\rm \bf PReB}$ as a collisional or repeated interaction model}
\label{Sec:collisional_models}
The PReB process may be thought of as a collisional or repeated-interaction model \cite{RauPR, ScaraniPRL, ZimanPRA, ciccarello2017, Campbell_2021}, where the system repeatedly interacts with multiple finite-sized chains. Collisional or repeated-interaction models have provided valuable insight in a diverse range of settings, with quantum thermodynamics~\cite{BarraSciRep, GabrieleNJP2018, StrasbergPRX, Guarnieri2020PhysLettA} and non-Markovian dynamics~\cite{CiccarelloPRA, VacchiniPRL, StrunzPRA, CampbellPRA, CompositeCMs, CompositeCMs2} being particularly elegant examples. Part of the appeal relies on their computational simplicity which, under suitable constraints, recovers well known dynamics captured by the Lindblad equation \cite{Campbell_2021}. They are often taken as the starting point or toy model for a particular purpose and are often limited to a single collisional unit, multipartite collisional models with several single-qubit baths being only recently explored \cite{Cattaneo_2021}. 

Despite their versatility, collisional models have been known to suffer some notable limitations. First, it has been unclear whether simple repeated interaction models, where the collisional unit consists of only a single or small number of constituents, can accurately represent a bath with a given spectral function. Second, with only few notable exceptions, it has been unclear whether the inherently discrete dynamics captured by a collision model could be exploited to extract results for continuous dynamics in presence of infinite baths beyond the Markovian regime.
PReB allows us to overcome both of these issues (via chain mapping for the former and via varying collision time for the latter), thus significantly extending the range of applicability of repeated interaction schemes. Further, it is worth stressing that rather than {\it a priori} starting from a collisional model, PReB arrives at one from the general considerations leading to Eq.~\eqref{PReB_statement0}.

\section{Numerical results}
\label{Sec:numerical_results}

\subsection{Results in fermionic chains}

For numerical demonstration of above discussion, we consider the following one-dimensional ordered interacting fermionic system,
\begin{align}
\label{XXZ}
\hat{\mathcal{H}}_S &= \sum_{\ell=1}^{L_S-1}\left( \hat{c}_\ell^\dagger \hat{c}_{\ell+1}+\hat{c}_{\ell+1}^\dagger \hat{c}_{\ell} + V \hat{n}_\ell \hat{n}_{\ell+1}\right)+h\sum_{\ell~{\rm odd}}\hat{n}_\ell.
\end{align}
where $L_S$ is the number of sites in the chain, $\hat{c}_\ell$ is the fermionic annihilation operator at site $\ell$ of the chain, $\hat{n}_\ell=\hat{c}_\ell^\dagger \hat{c}_\ell$, $V$ is the strength of nearest neighbour repulsive interaction, $h$ is strength of a potential that acts only on odd sites and we have set the hopping parameter to $1$. We consider a two-terminal set-up with two fermionic baths coupled at the first and last sites of the chain, 
\begin{align}
&\hat{\mathcal{H}}_{SB}^{(1)}=\sum_{r=1}^\infty\kappa_{r1}(\hat{c}_1^\dagger \hat{B}_{r1} + \hat{B}_{r1}^\dagger\hat{c}_1 ), \nonumber \\
&\hat{\mathcal{H}}_{SB}^{(2)}=\sum_{r=1}^\infty\kappa_{r2}(\hat{c}_{L_S}^\dagger \hat{B}_{r2} + \hat{B}_{r2}^\dagger\hat{c}_{L_S} ).
\end{align}
The baths are initially in thermal states with their own respective inverse temperatures $\beta_1$, $\beta_2$ and chemical potentials $\mu_1$, $\mu_2$. For simplicity, we assume the spectral functions of the baths to be of the form 
\begin{align}
\mathfrak{J}_\ell(\omega)=\Gamma_\ell\sqrt{1-\left(\frac{\omega}{2g_B}\right)^2},~~\Gamma_\ell=\frac{2\gamma_\ell^2}{g_B},~\ell=1,2.
\end{align}
After the chain-mapping, this corresponds to a non-interacting tight-binding chain with constant hopping parameter $g_B$, and zero on-site energies, while the hopping strength between system and bath is given by $\gamma_\ell$. To numerically simulate dynamics up to time $t$, we use baths of the size $L_B=(t+1)g_B$. The results are unaffected for larger $L_B$. 

Via a Jordan-Wigner transform, the Hamiltonian in Eq.(\ref{XXZ}) is exactly mapable to a Heisenberg XXZ-chain with a staggered field $h$. Without the staggered field, i.e, for $h=0$, it is a canonical model for an interacting integrable system in one-dimension \cite{Takahashi_book}. Its out-of-equilibrium properties in various regimes have remained of great interest both theoretically \cite{Bulchandani_2020,Bulchandani_2021,Znidaric_2011_XXZ,Ljubotina_2017,Bertini_2020_review,Bertini_2016,Castro_2016} and experimentally \cite{Jepsen_2021,Jepsen2020,Scheie2021,Chu_2020}. The presence of the staggered field breaks integrability, making it a model for generic systems \cite{PhysRevLett.125.180605}.  Obtaining dynamics of either the integrable or the non-integrable case, in the two-terminal open system set-up, which is directly relevant for applications in quantum heat-engines and refrigerators \cite{benenti2017fundamental}, has remained an outstanding problem beyond weak system-bath couplings \cite{Nathan_Rudner_2020}. To demonstrate that such outstanding problems become addressable in the PReB formalism, we choose strong unequal system-bath couplings ($\Gamma_1=1$, $\Gamma=2$), finite temperatures ($\beta_1=0.1$, $\beta_2=0.2$) and chemical potentials ($\mu_1=1.5$, $\mu_2=-1.5$), a finite but extended system ($L_S=16$). For these bath parameters, $\tau_M\sim 2$ is estimated as the time when the bath correlation functions decay below $5\%$ of their highest value.   We choose the initial state of the system as the half-filled product state.

\begin{figure}
\includegraphics[width=\columnwidth]{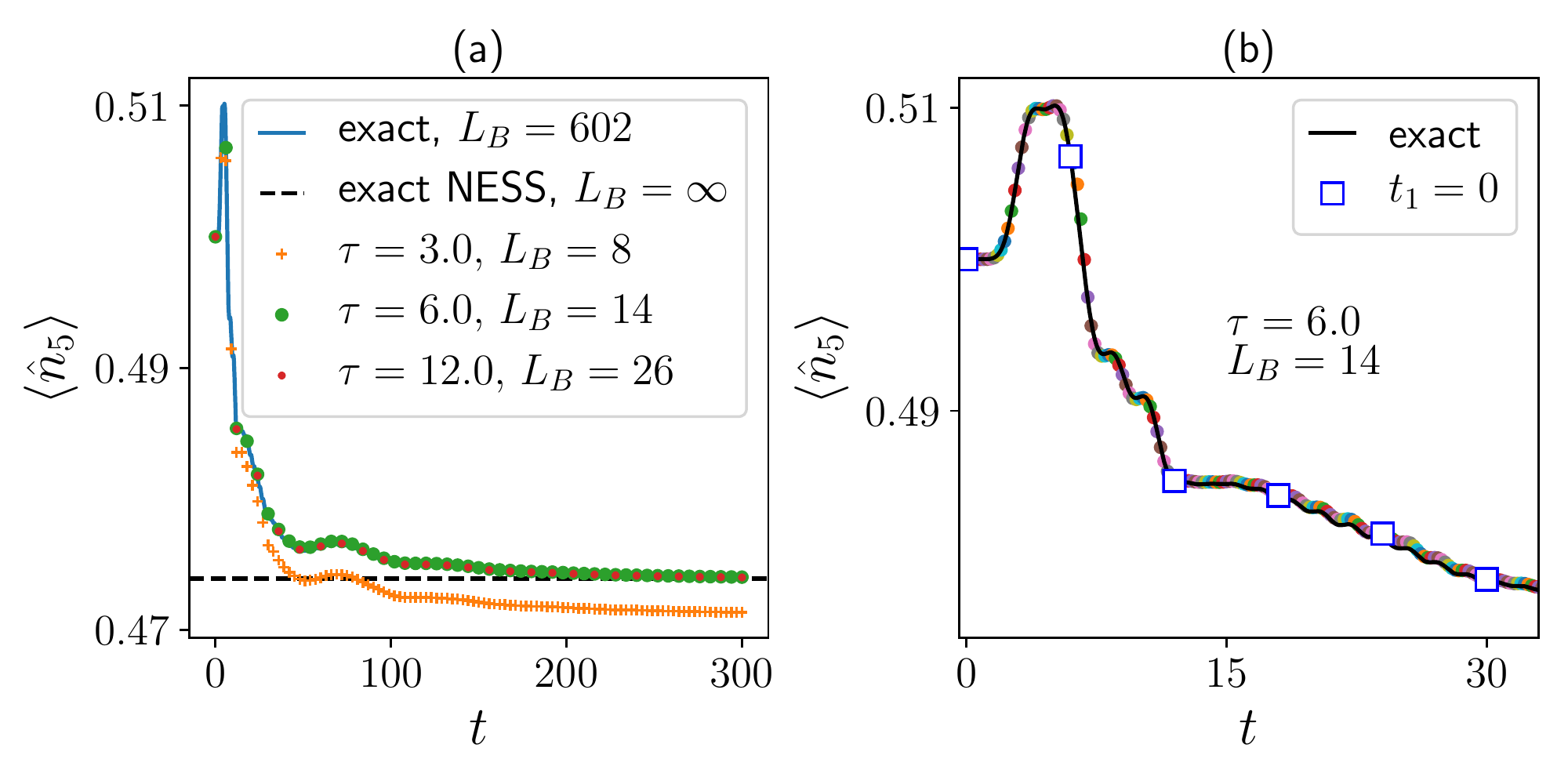} 
\caption{(a) Convergence to exact results with increasing the PReB time step $\tau$ is shown for a representative observable for the non-interacting chain ($V=0, h=0$). (b)~Reconstructing results at all times by repeating PReB simulation with $\tau=6$, for different values of $0< t_1 < \tau$. Each dot of same color is data obtained for one choice of  $t_1$. Parameters: $L_S=16$, $\beta_1=0.1$, $\beta_2=0.2$, $\mu_1=1.5$, $\mu_2=-1.5$, $g_B=2$, $\Gamma_1=1$, $\Gamma_2=2$. All energy scales are in units of system hopping parameter.  }
\label{fig:non_int} 
\end{figure}

\subsubsection{Benchmark: Non-interacting case: $V=0$, $h=0$}

We first consider the non-interacting case, $V=0$, which allows benchmarking against exact results. For simplicity, we will also set $h=0$ in this case. The exact dynamics can be obtained by rewriting the full system-bath Hamiltonian as $\hat{\mathcal{H}}=\sum_{\ell,m} \mathbf{H}_{\ell m} \hat{d}_\ell^\dagger \hat{d}_m$, where  $\ell, m$ refers to either system or bath sites, and numerically obtaining the correlation matrix $\mathbf{C}_{p q}(t) = {\rm Tr}\left( \hat{\rho}_{tot}(t) \hat{d}_p^\dagger \hat{d}_q \right)$ using $\mathbf{C}(t)=e^{i\mathbf{H} t} \mathbf{C}(0)e^{-i\mathbf{H} t}$. For comparison, non-equilibrium steady state (NESS) results with infinite baths are obtained exactly using the NEGF approach (see Appendix~\ref{app:NEGF} for explicit formulas). In Fig.~\ref{fig:non_int}(a) we show the convergence of PReB results for the non-interacting system with increasing $\tau$, for a representative observable.  We check for convergence with increasing $\tau$ by doubling its value. Such a convergence is clear in Fig.~\ref{fig:non_int}(a), where the results from $\tau=6$ and $\tau=12$ cases coincide. In contrast, for $\tau=3$ the results are clearly different, and hence is not converged. Furthermore, the converged results match with the exact dynamics in the long time limit, reaching the same NESS values. The converged PReB results in Fig.~~\ref{fig:non_int}(a) are obtained with simulations starting from the initial state. In terms of Eq.(\ref{PReB_statement0}), it means $t_1=t_0=0$. This gives results at time points which are multiples of $\tau$. Results at all other time points can be reconstructed by repeating the PReB simulation for same choice of $\tau$ but with different choices of $t_1$, $0<t_1<\tau$. This is shown in Fig.~~\ref{fig:non_int}(b) for a representative observable, where,  remarkably, we are able to reconstruct the full time evolution of the system in presence of the infinite thermal baths, by repeatedly using simulation up to $\tau=6$ requiring finite-sized baths, $L_B=14$, whose sizes are smaller than the system-size $L_S=16$.

\subsubsection{Interacting integrable case: $V=1$, $h=0$}

Next, we go to the interacting integrable case with $V=1$, $h=0$. In this case, we have to directly deal with time evolution of a mixed state of a chain of Hilbert space of dimension $2^{L_S+2 L_B}$, an extremely challenging problem, with complexity scaling exponentially with size of the chain. One standard way to obtain dynamics in such cases, in the absence of any small parameter, is by using tensor network time-evolution techniques for mixed states, a plethora of which exist \cite{Vidalmix,Verstraete_Cirac_2004,Feiguin_White_2005,Scholl05,
Stoudenmire_2010,Binder_Barthel_2015,tensor_network_review_2019,DMT}. The main approximation in all such techniques is convergence of dynamics with a finite bond-dimension, $\chi$. The converged dynamics so obtained is numerically exact. The complexity then becomes linear in size of the chain, but grows exponentially with the $\chi$ required for convergence. 
Crucially, in a unitary evolution, the $\chi$ required for convergence typically grows with time, making accurate long-time evolution (as required when $t_{\rm ss}$ is large) intractable. 

\begin{figure}
\includegraphics[width=\columnwidth]{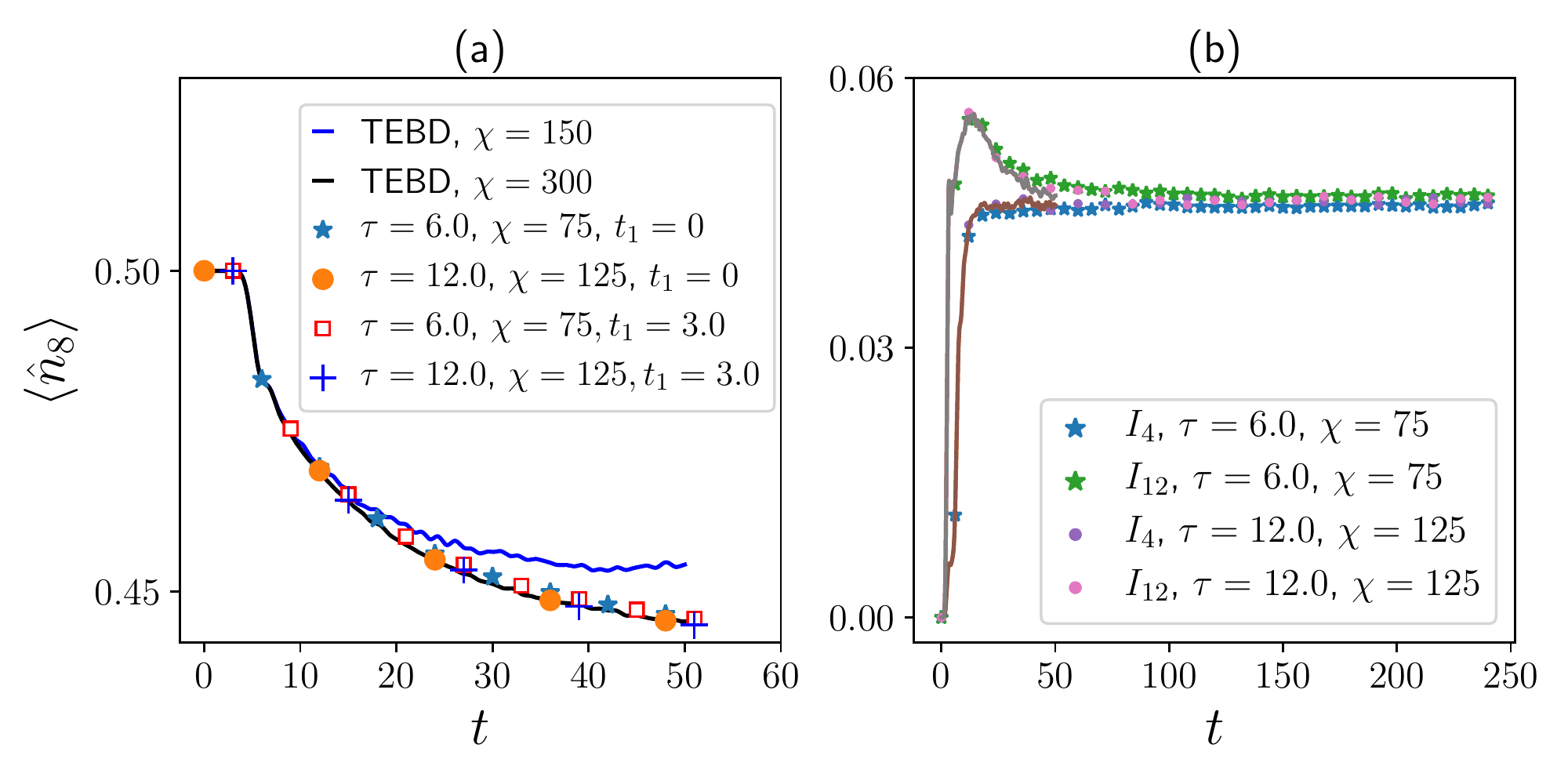} 
\caption{(a) Convergence of TEBD and PReB with TEBD approaches for the integrable interacting system ($V=1, h=0$) is shown for a representative observable. (d) The convergence of PReB with TEBD approach is shown up to long times for local particle currents  at two different sites of the interacting system. The continuous lines show TEBD results for $\chi=300$ up to $t=50$.  For continuous time evolution $L_B=102$. For PReB with TEBD, $L_B=14$ for $\tau=6$ and $L_B=26$ for $\tau=12$. Parameters: $L_S=16$, $\beta_1=0.1$, $\beta_2=0.2$, $\mu_1=1.5$, $\mu_2=-1.5$, $g_B=2$, $\Gamma_1=1$, $\Gamma_2=2$. The Trotter time step for TEBD is $0.1$. All energy scales are in units of system hopping parameter.  }
\label{fig:integrable} 
\end{figure}

It is exactly here that Eq.(\ref{PReB_statement0}) becomes extremely useful. This is demonstrated in Fig.\ref{fig:integrable}(a), where we plot the dynamics of a representative observable up to time $t=50$ as obtained from continuous time evolution and from PReB, using a mixed-basis \cite{Zwolak_2020,Wolf2014,mesoscopic_baths_fermions3} version of the standard time-evolution-by-block-decimation (TEBD) tensor network technique. The explicit details of this technique for our set-up is given in Appendix~\ref{app:TEBD_mixed_basis}. The continuous time evolution needs $\chi=300$ to converge up to $t=50$, while smaller values of $\chi$ converge up to smaller times, as shown for $\chi=150$. On the other hand, we find that PReB for $\tau=6$ ($L_B=14$) converged with $\chi=75$, and PReB for $\tau=12$ ($L_B=26$) converged with $\chi=125$. The PReB results have also converged with $\tau$, as the results for $\tau=6$ and $\tau=12$ match very well with each other, and with that from continuous time evolution. This is shown both for a choice of $t_1=t_0=0$ and $t_1=3>t_0$, thereby demonstrating that reconstruction of all time points is possible with PReB. To put the extreme numerical advantage of the PReB simulation into perspective, obtaining the continuous time TEBD result with $\chi=300$ up to $t=50$ required a wall-time of about $90$ hours, while obtaining the same with PReB required only about $12$ minutes for $\tau=6$, and about $1$ hour for $\tau=12$,  in the same computer architecture (Intel i9 10th Generation 8 core, 16 threads processor).

Furthermore, while continuous evolution up to longer times requires larger bond-dimensions rendering it intractable, continuing the PReB simulation for a chosen $\tau$ up to long times does not. This allows obtaining full time evolution up to steady state with a small bond-dimension using PReB, even when the same would not be possible with continuous time evolution. This is demonstrated in Fig.~\ref{fig:integrable}(d), where dynamics of local particle currents, defined as $I_{\ell} = 2i \langle \hat{c}^\dagger_{\ell+1} \hat{c}_\ell-  \hat{c}^\dagger_{\ell} \hat{c}_{\ell+1}\rangle$, at two different sites, as obtained from PReB with $\tau=6$ and $\tau=12$ are shown. At the final time-point, the currents at different bonds are almost the same, showing that the NESS has been approximately reached. Therefore, not only have we verified Eqs.(\ref{PReB_statement0}), (\ref{tau_condition0}) in a non-trivial system, as a consequence, we have been able to obtain numerically exact completely non-perturbative dynamics up to NESS of an interacting integrable quantum many-body system strongly connected to two baths which were initially at different finite temperatures and chemical potentials, a case which has previously remained intractable despite relevance in several contexts \cite{Bulchandani_2020,Bulchandani_2021,Znidaric_2011_XXZ,Ljubotina_2017,Bertini_2020_review,Bertini_2016,Castro_2016,Jepsen_2021, Jepsen2020,Scheie2021,Chu_2020}.  

\subsubsection{Non-integrable case: $V=1$, $h=1$}

\begin{figure}
\includegraphics[width=\columnwidth]{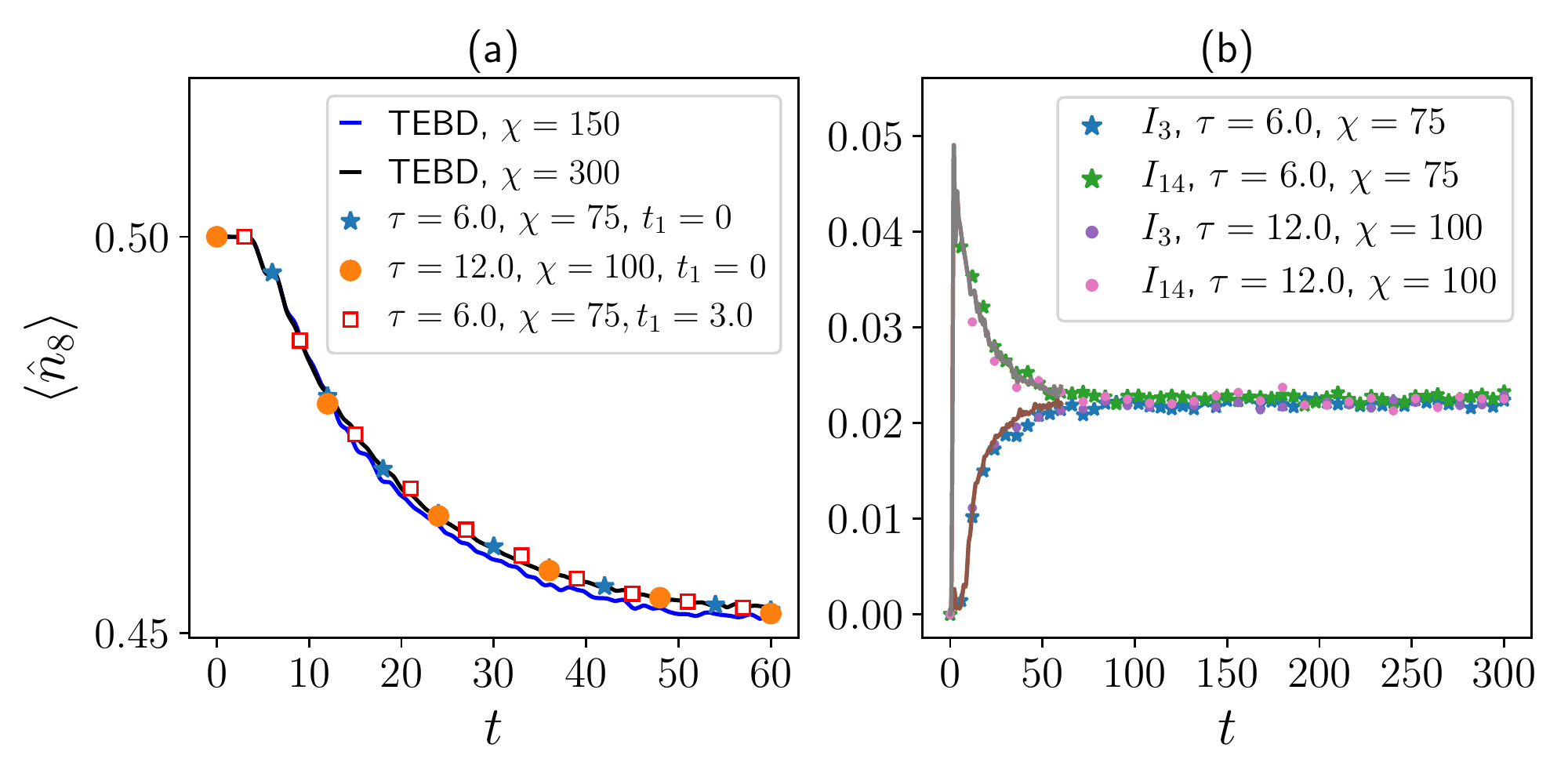}
\caption{(a) Convergence of TEBD and PReB with TEBD approaches for the non-integrable interacting system ($V~=~1$, $h~=~1$) with is shown for a representative observable. (b) The convergence of PReB with TEBD approach is shown up to long times for  local particle currents at two different sites for the non-integrable interacting system. The TEBD results for $\chi=300$ up to $t=60$ are also shown with the continuous lines. Parameters: $\beta_1=0.1$, $\beta_2=0.2$, $\mu_1=1.5$, $\mu_2=-1.5$, $g_B=2$, $\Gamma_1=1$, $\Gamma_2=2$. The Trotter time step for TEBD is $0.1$. All energy scales are in units of system hopping parameter.  }
\label{fig:staggered_plots} 
\end{figure}

Now, we look at the non-integrable case with $V=1$, $h=1$, using the same numerical technique. Note that, it is usually believed that in non-integrable systems, due to internal chaotic dynamics, the exact description of the baths matter less, while the same is not the case for integrable systems. In this sense, our results above for the interacting integrable system are more non-trivial than the results for the non-integrable system.

The results for the non-integrable system are shown in Fig.~\ref{fig:staggered_plots}(a), where we plot the dynamics of a representative observable up to time $t=60$ as obtained from continuous time evolution, from PReB with $\tau=6$ and $\tau=12$ with $t_1=0$, and from PReB with $\tau=6$ with $t_1=3$. It is clear that our main results Eqs.(\ref{PReB_statement0}), (\ref{tau_condition0})  are satisfied, and the PReB simulation requires much smaller bond dimensions. The relative computational resource and time advantages are of the same order as in the integrable case. However, both PReB results and the continuous evolution results seem to require a smaller bond-dimension to converge, as compared to the integrable case.  The long-time dynamics for local currents at two different sites, as obtained from the converged PReB process are shown in Fig.~\ref{fig:staggered_plots}(b). As before, the currents at various bonds become approximately same with increase in time, showing that NESS has been approximately reached.  The NESS seem to be reached at a shorter time in the non-integrable system than in the integrable case. This is likely due to the internal chaos of the system. This deserves to be investigated in more detail and will be taken up in future works.  We will like to stress that, both in the interacting integrable and non-integrable cases presented here, there is no small parameter in the Hamiltonian. All parameters, including the strength of system-bath couplings, are of the same order.



\subsection{Results in a spin-boson model using a completely different technique}

All numerical results till now have been for the two-terminal fermionic set-up. Further, both the tensor network technique used for interacting systems and the correlation matrix technique used for non-interacting systems, depend on chain-mapping of the baths. Here, to demonstrate that the PReB formalism is independent of such details, we check our main results, Eqs.(\ref{PReB_statement0}), (\ref{tau_condition0}), on a completely different system with a completely different technique. For this we choose one of the most canonical models of open quantum dynamics, a qubit  undergoing thermalization with a bosonic bath with Ohmic spectral density. The Hamiltonian of the set-up is given by $\hat{\mathcal{H}}=\hat{\mathcal{H}}_S+\hat{\mathcal{H}}_{SB}+\hat{\mathcal{H}}_B$,
\begin{align}
& \hat{\mathcal{H}}_S = \frac{\varepsilon}{2}\hat{\sigma}_z + \Delta \hat{\sigma}_x,  \\
& \hat{\mathcal{H}}_{SB}=\hat{\sigma}_z \sum_{r=1}^\infty \left( \kappa_{r} \hat{B}_r + \kappa_r^* \hat{B}_r^\dagger \right),
~ \hat{\mathcal{H}}_{B}=\sum_{r=1}^\infty \Omega_r \hat{B}_r^\dagger\hat{B}_r, \nonumber
\end{align} 
where $\hat{B}_r$ is now the annihilation operator for the $r$th bosonic mode of the bath. We choose the spectral density of the bath in the Ohmic form, with Gaussian cut-off
\begin{align}
\mathfrak{J}(\omega) = \gamma_b \omega e^{-\left(\omega/\omega_c\right)^2} \Theta(\omega),
\end{align}
where $\gamma_b$ gives the strength of coupling to bath, $\omega_c$ is the cut-off frequency, and $\Theta(\omega)$ is the Heaviside theta function. The bosonic bath is initially taken to be in a thermal state with inverse temperature $\beta$ (and chemical potential is zero).

Instead of chain-mapping techniques, we use a completely different path-integral based technique that has been utilized to numerically exactly solve such systems for a wide range of parameters. This numerical technique, called TEMPO (acronym for time-evolving-matrix-product-operator) \cite{TEMPO}, is based on exactly integrating out the infinite baths in the Feynman-Vernon influence functional approach. It utilizes the so called augmented-density-tensor \cite{Makri_1995_I,Makri_1995_II} to take care of the non-Markovian memory effects efficiently using tensor networks. Unlike chain-mapping techniques, TEMPO does not depend of the effective finiteness of baths, but instead crucially requires the baths to have infinite degrees of freedom.  Numerical code implementing TEMPO is available as a python package \cite{TEMPO_collab} and is particularly easy to use for spin-boson models.

\begin{figure}[t]
\includegraphics[width=\columnwidth]{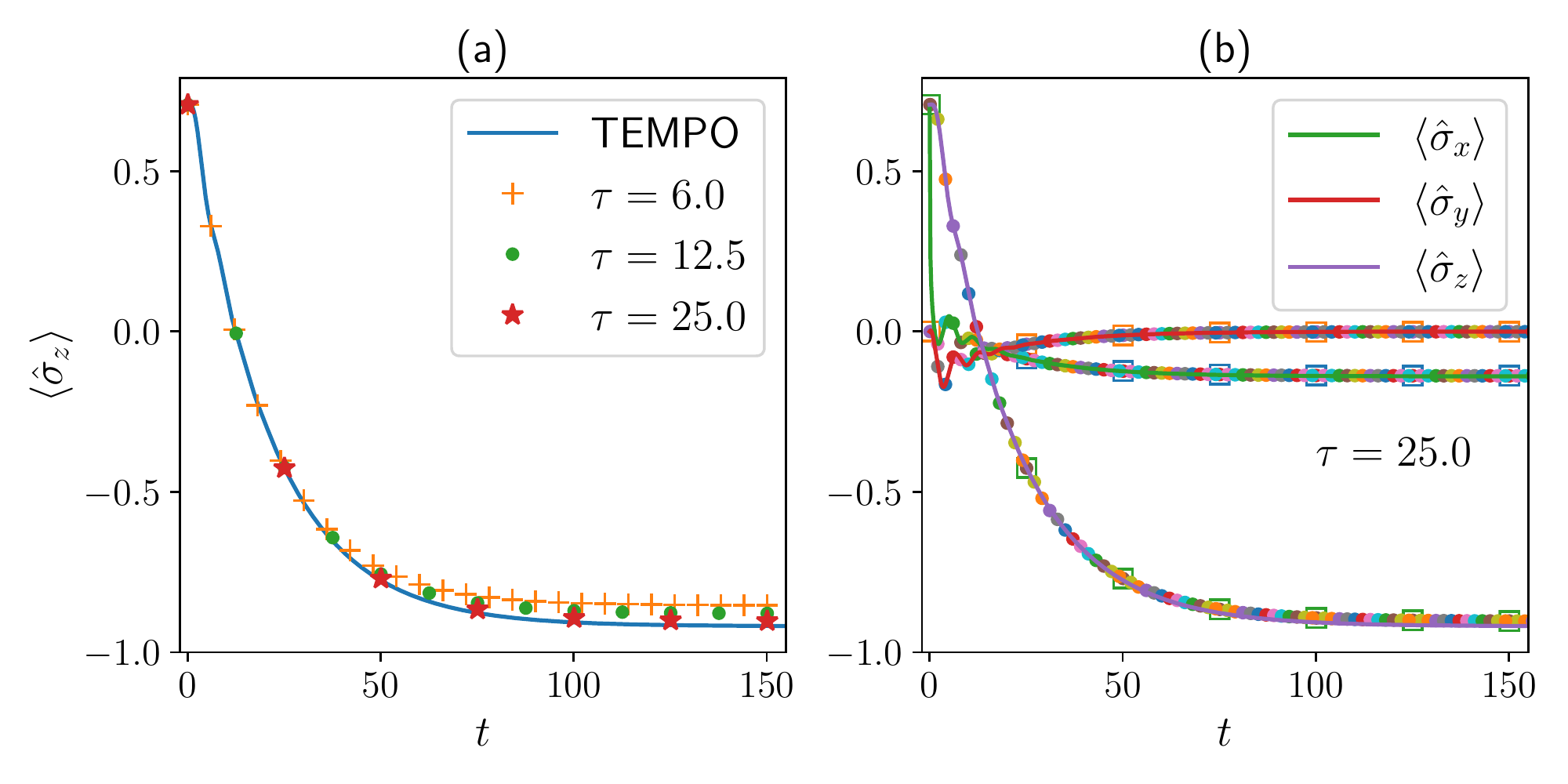}
\caption{(a) The plot shows dynamics of $\langle\hat{\sigma   }_z \rangle$ of a spin-boson model as obtained from numerically exact TEMPO technique, and as from PReB with TEMPO for different choices of $\tau$. The convergence of PReB results to the exact results is clear. (b) Reconstruction of full dynamics of the up to steady state by varying $t_1$ in regime $0\leq t_1<\tau$ is shown for PReB with $\tau=25$. The squares highlight the points for $t_1=0$.  Parameters: $\Delta=0.3$, $\gamma_b=0.1$, $\beta=10$, $\omega_c=50$. All energy scales are in units of $\varepsilon$.}  
\label{fig:TEMPO} 
\end{figure}

Validity of PReB for the spin-boson model with TEMPO is shown in Fig.~\ref{fig:TEMPO}. The plot corresponds to low temperature $\beta=10$ (all parameters in units of $\varepsilon$), and strong system-bath coupling strength $\gamma_b=0.1$. The memory time $\tau_M$ can be estimated to be $\tau_M \sim 3$, which rules out any possibility of a Markovian quantum master equation description. Fig.~\ref{fig:TEMPO}(a) shows convergence with increasing $\tau$, while Fig.~\ref{fig:TEMPO}(b) shows full reconstruction of dynamics by varying $t_1$. Clearly both Eqs.(\ref{PReB_statement0}), (\ref{tau_condition0}) are satisfied. 

TEMPO is one of the whole class of techniques based on storing memory effects \cite{Makri_1995_I,Makri_1995_II,TTM,TEMPO,process_tensor}. Scalability of these techniques to long-time simulation of extended systems is severely limited by the computational cost of storing the finite memory effects of an extended system up to long times. Thus, even in these cases, the ability to reconstruct the full dynamics by recursively using finite time evolutions will be useful, and possibly allow scaling these techniques to complex extended systems. 

If $\Delta=0$, the chosen model corresponds to the so called independent boson model and can be exactly solved. However, in this case, $\hat{\sigma}_z$ would be a conserved quantity of the global dynamics. This means that there would be no unique steady state, and so, the main assumption in derivation of PReB would not hold. Thus, for $\Delta=0$, the PReB approach would not work in general, although further investigations are required to make more concrete statements.

\section{Conclusions and Outlook}
\label{Sec:conclusions}

Thus, our PReB formalism, based on Eqs.(\ref{PReB_statement0}) and (\ref{tau_condition0}), significantly extends the classical simulability of open quantum systems, allowing for numerically exact simulations of non-Markovian dynamics of one-dimensional systems up to long times. Two-terminal transport through Hamiltonians of the form Eq.(\ref{XXZ}), with various additional on-site potentials, are ubiquitous in studies of quantum transport, localization and integrability-breaking \cite{Marko, Znidaric_2011_XXZ,mendoza2015coexistence, vznidarivc2016diffusive,vznidarivc2017dephasing,PhysRevB.99.094435,schulz2020phenomenology,vznidarivc2018interaction,PhysRevB.100.085105,brenes2018high,PhysRevLett.125.180605}. But they have been limited to the infinite temperature Markovian regime, while other regimes have remained intractable.  The PReB formalism enables a whole class of existing numerical techniques \cite{Makri_1995_I,Makri_1995_II,TEMPO,process_tensor,thermofield_bosons,
TEDOPA_bosons,TEDOPA_fermions,Schwarz_2018,Boulat_2008,Tamascelli_PRL} to access to these previously intractable regimes, which is important for a wide range of applications \cite{goold2016role,benenti2017fundamental, datta1997electronic,akkermans2007mesoscopic, lambert2013quantum,Molecular_junction_review,deVega_2017_review}. Moreover, since the formalism is general, it may allow adaptation of these techniques to higher dimensional open systems, as well as to generic baths.  At a more fundamental level, Eq.(\ref{PReB_statement0}) points to a deep connection between Markovian and non-Markovian dynamics in open quantum systems. Due to its direct relation with collisional or repeated interaction models, PReB also opens the possibility to significantly extend the understanding established in these models \cite{RauPR, ScaraniPRL, ZimanPRA, ciccarello2017, Campbell_2021,BarraSciRep, GabrieleNJP2018, StrasbergPRX, Guarnieri2020PhysLettA,CiccarelloPRA, VacchiniPRL, StrunzPRA, CampbellPRA, CompositeCMs, CompositeCMs2,Cattaneo_2021}. Going further, our results may allow novel ways of bath-engineering to obtain target steady states. These directions will be investigated in future works.

\paragraph*{Acknowledgements.--}
We acknowledge support from the European Research Council Starting Grant ODYSSEY (G. A. 758403), the SFI-Royal Society University Research Fellowship scheme. SC acknowledges the Science Foundation Ireland Starting Investigator Research Grant ``SpeedDemon" (No. 18/SIRG/5508). AP acknowledges funding from European Unions Horizon 2020 research and innovation programme under the
Marie Sklodowska-Curie grant agreement No. 890884.
AP acknowledges Irish Centre for High End Computing
(ICHEC) for the provision of computational facilities. GG acknowledges support from FQXi grant (DFG FOR2724).  J.P. is grateful for financial support from Ministerio de Ciencia, Innovaci{\'o}n y Universidades (SPAIN), including FEDER (Grant Nos. PGC2018-097328-B-100) together with Fundaci{\'o}n S{\'e}neca (Murcia, Spain) (Project No. 19882/GERM/15). AP thanks Gerald Fux for useful discussions. We also thank the anonymous referee for extremely insightful suggestions which helped improve the paper manifold.

\section*{Appendix}

\appendix

\section{Derivation of Periodically Refreshed Baths}\label{app:derivation}

\subsection{Set-up}
We consider the general set-up of a system connected to (possibly multiple) baths, starting from a product state of the system and the baths, and evolving according to a global system+baths Hamiltonian. The dynamics of the system is given by
\begin{align}
\label{def_Lambda}
&\hat{\rho}(t)= \hat{\Lambda}(t-t_0)[\rho(t_0)] = {\rm Tr}_B(\hat{\rho}_{tot}(t)) \nonumber \\
& \hat{\rho}_{tot}(t) = {\rm Tr}_B\Big(e^{-i\hat{\mathcal{H}}(t-t_0)}\hat{\rho}(t_0) \hat{\rho}_B e^{i\hat{\mathcal{H}}(t-t_0)} \Big), 
\end{align} 
where $\rho_B$ is the initial state of the baths, $\hat{\rho}(t_0)$ is the initial state of the sytem, and 
\begin{align}
\hat{\mathcal{H}}= \hat{\mathcal{H}}_S + \hat{\mathcal{H}}_{SB} + \hat{\mathcal{H}}_B,
\end{align}
is the full Hamiltonian of the set-up, $\hat{\mathcal{H}}_S$ being the system Hamiltonian, $\hat{\mathcal{H}}_{SB}$ being the system-baths coupling Hamiltonian, and 
$\hat{\mathcal{H}}_B$ being the Hamiltonian of the baths. The superoperator $\hat{\Lambda}(t-t_0)$ describes a completely positive trace preserving (CPTP) map which maps $\hat{\rho}(t_0)$ to $\hat{\rho}(t)$. Such a map is a universal dynamical map in the sense that $\hat{\Lambda}(t-t_0)$ is independent of the initial state $\hat{\rho}(t_0)$. Without loss of generality, we can assume ${\rm Tr}_B(\hat{\mathcal{H}}_{SB} \hat{\rho}_B)=0$. If this is not the case, the system Hamiltonian can always be slightly modified to satisfy this.  Further, in complete generality, we can assume that time has been rescaled such that it can be considered a dimensionless parameter.

In complete generality, the equation governing the time-evolution of $\hat{\rho}(t)$ can be written as \cite{Breuer_book}
\begin{align}
\label{NZ1}
\frac{\partial \hat{P}\hat{\rho}_{tot}}{\partial t} &=i[\hat{P}\hat{\rho}_{tot}(t),\hat{\mathcal{H}}_S] \nonumber \\
&+ \hat{P}\hat{\mathcal{L}} \int_0^{t-t_0} dt^\prime e^{t^\prime \hat{Q} \hat{\mathcal{L}}} \hat{Q}\hat{\mathcal{L}} \hat{P}\hat{\rho}_{tot}(t-t^\prime),
\end{align}
where the superoperators $\hat{P}$, $\hat{Q}$ and $\hat{\mathcal{L}}$ are defined as
\begin{align}
\label{def_PQL}
& \hat{P}(\bullet)=Tr_B(\bullet)\rho_B,~~\hat{Q}=\hat{\mathbb{I}}-\hat{P}, \\
& \hat{\mathcal{L}}(\bullet) = i [\bullet, \hat{\mathcal{H}}],
\end{align}
$\hat{\mathbb{I}}$ being the identity superoperator. After a bit of algebra, Eq.(\ref{NZ1}) can be written in the form
\begin{align}
\label{memory_kernel}
&\frac{\partial \hat{\rho}}{\partial t} = i [\rho(t), \hat{\mathcal{H}}_S]+ \int_{0}^{t-t_0} dt^\prime \hat{K}(t^\prime)[\hat{\rho}(t-t^\prime)], \nonumber \\
& \hat{K}(t)[\bullet]={\rm Tr_B}\left( \hat{\mathcal{L}}e^{t^\prime \hat{Q} \hat{\mathcal{L}}}\hat{Q}\hat{\mathcal{L}}\hat{P}[\bullet] \right)
\end{align}
Here $\hat{K}(t)$ is called the memory kernel superoperator. Further,  the time evolution can also be formally cast into the following equivalent, but apparently time-local, form, \cite{Chruscinski_2010}
\begin{align}
\label{time_local}
&\frac{\partial \hat{\rho}}{\partial t} = \hat{L}(t-t_0)[\hat{\rho}(t)], \nonumber \\
&\hat{L}(t)=\frac{d}{dt}[\hat{\Lambda}(t)] \hat{\Lambda}^{-1}(t),
\end{align}
where $\hat{\Lambda}^{-1}(t)$ is the formal inverse of $\hat{\Lambda}(t)$, i.e, $\hat{\Lambda}(t)\hat{\Lambda}^{-1}(t)=\mathbb{I}$. 
In the above form, $\hat{L}(t-t_0)$ is the generator of time evolution. The density matrix at time $t$ can be written as
\begin{align}
\hat{\rho}(t) = \mathcal{T}e^{\int_0^{t-t_0} dt^\prime \hat{L}(t^\prime) }\hat{\rho}(t_0),
\end{align}
where $\mathcal{T}$ represents time ordering. Yet another way to formally write down the exact equation governing the time-evolution of $\hat{\rho}(t)$ is \cite{Breuer_book}, 
\begin{align}
\label{memory_kernel2}
\frac{\partial\hat{\rho}}{\partial t}& = i[\hat{\rho}(t), \hat{\mathcal{H}}_S] + \int_{0}^{t-t_1} dt^\prime \hat{K}(t^\prime)[\hat{\rho}(t-t^\prime)] \nonumber \\
& +{\rm Tr_B}\left(\hat{\mathcal{L}}e^{(t-t_1) \hat{Q} \hat{\mathcal{L}}} \hat{Q}\hat{\rho}(t_1)\right),~~t_1>t_0.
\end{align}
Here, $\hat{K}(t)$ is the same memory kernel superoperator as in Eq.(\ref{memory_kernel}). The last term in above equation appears due to the fact that for $t_1>t_0$, the system and the bath are no longer in a product state.

Eqs.(\ref{NZ1}), (\ref{memory_kernel}), (\ref{time_local}), (\ref{memory_kernel2})  all describe the same dynamics, that given by Eq.(\ref{def_Lambda}), without any approximations. Given this setting, and armed with these equations, we now discuss the assumptions.

\subsection{Assumptions}

{\it Assumption 1: analyticity of dynamics ---}
The dynamics of the global set-up corresponds to that of a quench where the system-bath couplings are switched on at time $t_0$. We will assume that the dynamics following this quench remains analytic at all times. 

{\it Assumption 2: finite system Hilbert space dimension ---}
For technical reasons, we need to assume that the system Hilbert space is finite-dimensional. Strictly, this restricts us to fermionic or spin systems, and rules out bosons. Note that, this restriction is not there in the bath Hilbert space.  Further, even if the system is a lattice of bosonic sites, but there is some effective cut-off on the number of bosons at each site, either due to repulsive interactions, or due to temperatures in the problem, it amounts to an effective finite system Hilbert space dimension. Thus, this assumption is quite mild.

{\it Assumption 3: Unique steady state ---} This is the main assumption. It states that the long time state of the system is independent of the initial state. Many experimental situations of interest fall under this class. Formally, this means,
\begin{align}
\hat{\rho}_{\rm NESS}=\lim_{t_0 \rightarrow -\infty} \hat{\Lambda}(t-t_0)[\hat{\rho}(t_0)],
\end{align}
is unique, independent of $\hat{\rho}(t_0)$. In the language of CPTP maps, this means that one eigenvalue of $\hat{\Lambda}(t-t_0)$ will be $1$, while all other eigenvalues decay to zero with time. On physical grounds, this requires that the system-size is finite, while the baths are in the thermodynamic limit.

In the following, we discuss the consequences of these approximations.

\subsection{Consequences of the assumptions}

\subsubsection{Finite memory}
On physical grounds, it is clear that unique steady state implies a finite memory. To see this more rigorously, note that, the  steady state must be a solution of 
\begin{align}
0= i [\hat{\rho}_{\rm NESS}, \hat{\mathcal{H}}_S]+ \int_{0}^{\infty} dt^\prime \hat{K}(t^\prime)[\hat{\rho}(t-t^\prime)].
\end{align}
The right-hand-side of above equation explicitly depends on the initial state. This cannot be the case if the steady state is unique. So, the only way a unique steady state can be approached, is if, there exists a time $\tau_M$, such that,
\begin{align}
\label{finite_memory}
\Big| \Big|\int_{\tau_M}^{t-t_0} dt^\prime \hat{K}(t^\prime)[\rho(t-t^\prime)] \Big | \Big| <\epsilon~~\forall~~ t-t_0 \geq \tau_M,
\end{align}
where $| | \hat{O} | |$ refers the operator norm of $\hat{O}$ and $\epsilon$ is an arbitrarily small number set by numerical or experimental precision. This physically means that the memory kernel approximately decays to zero after a time $\tau_M$. Under this condition, we have,
\begin{align}
\label{finite_memory_QME}
\frac{\partial \hat{\rho}}{\partial t} \approx i [\hat{\rho}(t), \hat{\mathcal{H}}_S]+ \int_{0}^{\tau_M} dt^\prime \hat{K}(t^\prime)[\hat{\rho}(t-t^\prime)],~\forall~t-t_0 \geq\tau_M.
\end{align}
Now, the right-hand side of above equation does not depend explicitly on the initial state $\hat{\rho}(t_0)$, which has to be the case if $\hat{\rho}_{\rm NESS}$ is unique. Thus, Eq.(\ref{finite_memory}) is a necessary (not sufficient) condition for the steady state to be unique.

Writing Eq.(\ref{finite_memory_QME}) in the form of Eq.(\ref{time_local}), we get,
\begin{align}
\frac{\partial \hat{\rho}}{\partial t} \approx \hat{L}[\hat{\rho}(t)],~\forall~t-t_0 \geq\tau_M~~~\textrm{with }\hat{L}=\hat{L}(\tau_M).
\end{align}
Thus, the generator becomes approximately time independent after a time $\tau_M$. The solution of above equation is
\begin{align}
\label{finite_memory_rho1}
\hat{\rho}(t) \approx e^{(t-\tau_M) \hat{L}} [\hat{\rho}(t_0+\tau_M)],~~\forall~~t\geq t_0 + \tau_M
\end{align}
Let us define a superoperator $\hat{\mathcal{G}}$ which satisfies
\begin{align}
e^{\hat{\mathcal{G}}}[\hat{\rho}(t_0)]=\hat{\Lambda}(\tau_M)[\hat{\rho}(t_0)].
\end{align}
Since the dynamics remains analytic across the time $\tau_M$, $\hat{L}$ and $\hat{G}$ must commute,
\begin{align}
[\hat{L},\hat{\mathcal{G}}] = 0.
\end{align} 
So, we have,
\begin{align}
\label{finite_memory_rho2}
\hat{\rho}(t)=\hat{\Lambda}(t-t_0)[\hat{\rho}(t_0)] \approx e^{(t-\tau_M) \hat{L}+\hat{\mathcal{G}}} [\hat{\rho}(t_0)],~~\forall~t\geq t_0 +\tau_M.
\end{align}
Let $\{\lambda_i\}$ and $\{ g_i \}$ denote the eigenvalues of $\hat{L}$ and $\hat{\mathcal{G}}$ respectively, labelled in ascending order according to the magnitude of the real part of $\{\lambda_i\}$. Then, the uniqueness of steady state demands,
\begin{align}
\lambda_1=0,~~g_1=0,~~{\rm Re}(\lambda_i)<0,~~\forall~~i>1.
\end{align}
Let $t_{\rm ss}$ be the effective time to reach steady state, defined as
\begin{align}
|| \hat{\rho}(t) - \hat{\rho}_{\rm NESS} || < \epsilon,~~\forall~~t \geq t_{\rm ss}.
\end{align}
It follows from above that
\begin{align}
\label{t_ss}
t_{\rm ss} \gg \left|\frac{1}{{\rm Re}(\lambda_2)}\right|.
\end{align} 
Next, let us look at the effect of the assumptions on Eq.(\ref{memory_kernel2}). From Eq.(\ref{finite_memory}), for $t\geq t_1+t_0$, Eq.(\ref{memory_kernel2}) becomes
\begin{align}
\frac{\partial\hat{\rho}}{\partial t}& \approx i[\hat{\rho}(t), \hat{\mathcal{H}}_S] + \int_{0}^{\tau_M} dt^\prime \hat{K}(t^\prime)[\hat{\rho}(t-t^\prime)] \nonumber \\
& +{\rm Tr_B}\left(\hat{\mathcal{L}}e^{(t-t_1) \hat{Q} \hat{\mathcal{L}}} \hat{Q}\hat{\rho}(t_1)\right),~~t_1>t_0.
\end{align}
Since both the above equation and Eq.(\ref{finite_memory_QME}) describe the exact same process, we must have
\begin{align}
\Big| \Big| {\rm Tr_B}\left(\hat{\mathcal{L}}e^{(t-t_1) \hat{Q} \hat{\mathcal{L}}} \hat{Q}\hat{\rho}(t_1)\right) \Big| \Big|<\epsilon,~~\forall~~t-t_1\geq \tau_M.
\end{align}
 So, we get,
\begin{align}
\label{finite_memory_rho3}
\hat{\rho}(t) \approx e^{(t-\tau_M) \hat{L}} [\hat{\rho}(t_1+\tau_M)],~~\forall~~t\geq t_1 + \tau_M,~t_1 \geq t_0.
\end{align}
We use these results to derive the periodically refreshed baths in the following.

\subsubsection{Periodically Refreshed Baths}

Note, in Eq.(\ref{finite_memory_rho3}), that at time $t_1+\tau_M$, there are system-bath correlations. So the map generated by $\hat{L}$ is not a universal dynamical CPTP map. It is not guaranteed to map every given density matrix of the system to a density matrix at all times. But, it is guaranteed to map to density matrices those system density matrices that can be generated by time evolving the full set-up to time $t_1 + \tau_M$. Further, it is guaranteed to map every given density matrix to the steady state density matrix in the long-time limit. This means, we can always find a large enough value of $\tau$ such that
\begin{align}
\label{tau_def}
&\Big| \Big| e^{(t-\tau_M) \hat{L}} [\hat{\rho}(t_1+\tau_M)] - e^{(t-t_1) \hat{L}} [\hat{\rho}(t_1)] \Big| \Big| < \epsilon, \nonumber \\
& \forall~~t>t_1+\tau,~~\tau>t_1 \geq t_0.
\end{align}
The above condition is trivially satisfied if $\tau=t_{\rm ss}$. However, this needs not always be the case. To see this, we write the above condition for $t_1=t_0$, and use Eq.(\ref{finite_memory_rho2}), to check that $\tau$ is required to satisfy the following conditions
\begin{align}
\label{tau_condition}
& \tau \gg \tau_M-t_0 + \left | \frac{{\rm Re}(g_i)}{{\rm Re}(\lambda_i)} \right |,~~\forall~~i>1, \nonumber \\
& \tau \gg \tau_M-t_0 + \left | \frac{{\rm Im}(g_i)}{{\rm Im}(\lambda_i)} \right |,~~\forall~{\rm Im}(\lambda_i)>0, \\
& \tau \gg \left|\frac{1}{\lambda_i}\right |,~~\forall~{\rm Im}(\lambda_i)=0, ~{\rm Im}(g_i)\neq 0. \nonumber 
\end{align}
These set of conditions is not the same as the condition for $t_{\rm ss}$ in Eq.(\ref{t_ss}), unless ${\rm Im}(\lambda_2)=0$, ${\rm Im}(g_2)\neq 0$. Generically, we expect $\tau<t_{\rm ss}$, while in general, $t_{\rm ss}\geq \tau \gg \tau_M$.

From Eq.(\ref{finite_memory_rho2}) and Eq.(\ref{tau_def}), we have
\begin{align}
\hat{\rho}(t) \approx e^{(t-t_0) \hat{L}}[\hat{\rho}(t_0)],~~\forall~t\geq t_0 +\tau.
\end{align}
Noting that $\hat{\rho}(t)=\hat{\Lambda}(t-t_0)[\hat{\rho}(t_0)]$ and the initial state is arbitrary, we can write,
\begin{align}
\label{Lambda_L_mapping}
\hat{\Lambda}(t-t_0)[\bullet] \approx e^{(t-t_0) \hat{L}}[\bullet],~~\forall~t\geq t_0 +\tau.
\end{align}
Thus, the map generated by $\hat{L}$ becomes approximately the original universal dynamical map after a time $\tau$. From Eq.(\ref{finite_memory_rho3}) and Eq.(\ref{tau_def}), we get
\begin{align}
\hat{\rho}(t) \approx e^{(t-t_1)\hat{L}}[\hat{\rho}(t_1)],~~\forall~t\geq t_1+\tau,~t_1\geq t_0.
\end{align}
Choosing $t=n\tau+t_1$, we immediately see, using Eq.(\ref{Lambda_L_mapping}), that,
\begin{align}
&\hat{\Lambda}(n\tau+t_1-t_0)[\hat{\rho}(t_0)]\approx \underbrace{\hat{\Lambda}(\tau) [\ldots[\hat{\Lambda}(\tau)[}_{\text{n times}}\hat{\rho}(t_1)]]]\ldots].
\end{align}
The physical process described in the right-hand-side of the above equation corresponds to evolving up to time $t_1$, then detaching the baths and refreshing them to their original initial state, and afterwards periodically detaching the baths and refreshing them to their original initial state in steps to time $\tau$. The above equation says that the state of the system obtained from this process is approximately the same as the one obtained from continuous time evolution without any refreshing of baths up to time $n\tau+t_1$. It is clear from the definition of $\tau$ in Eq.(\ref{tau_def}), that this approximation becomes more and more accurate as $\tau$ is increased. This brings us to the main statement of periodically refreshed baths (PReB), which can be summarized as
\begin{align}
\label{PReB_statement}
&\hat{\rho}_{n\tau+t_1} = \underbrace{\hat{\Lambda}(\tau) [\ldots[\hat{\Lambda}(\tau)[}_{\text{n times}}\hat{\Lambda}(t_1-t_0)[\hat{\rho}(t_0)]]]\ldots]\nonumber \\
& \Big | \Big | \hat{\rho}(n\tau+t_1) - \hat{\rho}_{n\tau+t_1} \Big | \Big | = \epsilon(\tau), \textrm{ $\epsilon(\tau)$ decays with $\tau$.} 
\end{align}
In other words, $\hat{\rho}_{n\tau+t_1}$ converges to $\hat{\rho}(n\tau+t_1)$ with increase in $\tau$. The only assumptions required for this statement are the ones given in the previous section. Note that, in complete generality, we can restrict $t_1$ to $\tau>t_1\geq t_0$ to construct all time points. As argued in the main text, when our set-up describes a quantum many-body system on a lattice connected to multiple baths at few sites, we expect to find convergence of $\hat{\rho}_{n\tau+t_1}$ with a $\tau$ satisfying 
\begin{align}
\label{tau_condtion2}
t_{\rm ss} \gg \tau \gg \tau_M.
\end{align}
It is in these cases that Eq.(\ref{PReB_statement}) becomes extremely useful because, it says, long time evolution in presence of the baths can be reconstructed by repeatedly using simulation of much shorter time evolution.



\section{Estimating memory time}\label{app:estimating_memory}
It is clear from above that the value of $\tau$ required for convergence in PReB crucially depends on the effective memory time $\tau_M$ of the open system dynamics. Here we discuss how this time can be estimated directly from bath properties, without any reference to the system.

\subsection{Gaussian baths with general system-bath coupling}
The set-up we consider is governed by the full system+bath Hamiltonian $\hat{\mathcal{H}}=\hat{\mathcal{H}}_S+\sum_\ell \left(\hat{\mathcal{H}}_{SB}^{(\ell)}+\hat{\mathcal{H}}_B^{(\ell)}\right)$,
\begin{align}
\label{general_set_up}
\hat{\mathcal{H}}_B^{(\ell)} =    \sum_{r=1}^{\infty} \Omega_{r\ell} \hat{B}_{r\ell}^\dagger \hat{B}_{r\ell},~~\hat{\mathcal{H}}_{SB}^{(\ell)}=\sum_{\alpha, \ell}\hat{X}_{\alpha\ell} \hat{\mathcal{B}}_{\alpha\ell},
\end{align}
where  $\hat{B}_{r\ell}$ is the fermionic or bosonic annihilation operator of the $r$th mode of the $\ell$th bath, $\hat{X}_\alpha$ is a Hermitian operator of the system, and $\hat{\mathcal{B}}_\alpha$ is a bath operator. We assume the system Hilbert space is finite, so that $\hat{X}_\alpha$ has a finite spectral norm. At initial time, $t=t_0$, the system is assumed to be in an arbitrary state, uncoupled with the baths, while
 the baths are in thermal states with their individual temperatures and chemical potentials. Thus, the initial state of the whole set-up is given by
\begin{align}
\hat{\rho}_{tot}(t_0) = \hat{\rho}(t_0)\hat{\rho}_B,~~\hat{\rho}_B=\prod_{\ell}\frac{e^{-\beta_\ell(\hat{\mathcal{H}}_B^{(\ell)}-\mu_\ell \hat{N}_B^{\ell}})}{Z_B^{(\ell)}},
\end{align}
where $\hat{N}_B^{\ell}$ is the total number operator of the $\ell$th bath, $Z_B^{(\ell)}$ is the corresponding partition function, $\hat{\rho}(t_0)$ is an arbitrary initial state of the system. The coupling between the system and bath is switched on at time $t=t_0$ and the whole system+bath is then evolved unitarily to some time $t$ under the full set-up Hamiltonian $\hat{\mathcal{H}}$. Going to interaction picture with respect to $\hat{\mathcal{H}}_S$ and $\hat{\mathcal{H}}_B$, we have
\begin{align}
\hat{\rho}^I(t) = \hat{\Lambda}(t-t_0)[\hat{\rho}^I(t_0)]= {\rm Tr}_B \left(\hat{U}(t,t_0) \hat{\rho}^I(t_0)\hat{\rho}_B \hat{U}^\dagger(t,t_0)  \right)
\end{align}
$\hat{U}(t,t_0)=\mathcal{T}{\rm exp}\left(-i\int_{t_0}^t ds \hat{\mathcal{H}}_{SB}^I(s)\right)$, $\hat{O}^I(t)=e^{i(\hat{\mathcal{H}}_{S}+\hat{\mathcal{H}}_{B})t}\hat{O}(t)e^{-i(\hat{\mathcal{H}}_{S}+\hat{\mathcal{H}}_{B})t}$ for any operator $\hat{O}$, $\mathcal{T}$ denotes time-ordering and ${\rm Tr}_B \left(...\right)$ denotes trace over bath degrees of freedom. This is the standard microscopic approach to open quantum dynamics.

The exact quantum master equation for the above set-up was derived in Appendix A of Ref.\citep{Nathan_Rudner_2020}.  The exact quantum master equation is given in interaction picture by
\begin{align}
\label{exactQME_explicit}
\frac{\partial \hat{\rho}^I}{\partial t}=\int_{0}^{t-t_0} dt_1 \sum_{\alpha,\nu,\ell}  \Big(& Q_{\alpha \nu}^{(\ell)}(t_1)\left[ \hat{A}_{\nu \ell}(t,t-t_1),\hat{X}_{\alpha \ell}^I(t)\right]  \nonumber \\
& {+\rm h.c} \Big),
\end{align}
with $\hat{A}_{\nu}(t,t_1) = {\rm Tr}_B\left(\hat{U}(t,t_1)\hat{X}^I_{\nu \ell}(t_1)\hat{\rho}_{tot}^I(t_1)\hat{U}^{\dagger}(t,t_1)\right)$, h.c. denotes Hermitian conjugate and 
\begin{align}
Q_{\alpha \nu}^{(\ell)}(t) = {\rm Tr}_B\left(\hat{\rho}_B\hat{\mathcal{B}}_{\alpha\ell}^I(t)\hat{\mathcal{B}}_{\nu\ell}\right). 
\end{align}
Eq.(\ref{exactQME_explicit}) is same as Eq(A12) of Ref.\cite{Nathan_Rudner_2020}.
On the other hand, using Nakajima-Zwanzig projection operator method for our set-up leads to an equation of the form
\begin{align}
\frac{\partial \hat{\rho}^I}{\partial t}=\int_{0}^{t-t_0} dt_1 \hat{K}(t_1)[\hat{\rho}^I(t-t_1)],
\end{align}
which on comparing with Eq.(\ref{exactQME_explicit}) lets us identify
\begin{align}
\hat{K}(t_1)[\hat{\rho}^I(t-t_1)] = \sum_{\alpha,\nu,\ell}\Big(& Q_{\alpha \nu}^{(\ell)}(t_1)\left[ \hat{A}_{\nu \ell}(t,t-t_1),\hat{X}_{\alpha \ell}^I(t)\right]  \nonumber \\
& {+\rm h.c} \Big). 
\end{align}
Therefore, it is clear that the time for decays of $Q_{\alpha \nu}^{(\ell)}(t)$ gives the memory time $\tau_M$.  The error due to choosing a finite $\tau_M$ can be rigorously bounded. Splitting the time integration into two parts, one from $0$ to $\tau$ and another from $\tau_M$ to $t-t_0$, directly gives the expression for $\hat{\mathcal{E}}(t,t_0,\tau_M)$, which represents the terms that are neglected in making the approximation, as 
\begin{align}
&\hat{\mathcal{E}}(t,t_0,\tau_M)& \nonumber \\
&=\int_{\tau_M}^{t-t_0} dt_1 \sum_{\alpha,\nu,\ell}  \Big(
Q_{\alpha \nu}^{(\ell)}(t_1)\left[ \hat{A}_{\nu \ell}(t,t-t_1),\hat{X}_{\alpha \ell}^I(t)\right] {+\rm h.c} \Big).
\end{align}
Using results and techniques from Appendix A of Ref.\cite{Nathan_Rudner_2020}, it can be shown that,
\begin{align}
|| \hat{A}_{\nu}(t,t_1) || \leq || \hat{X}_{\nu \ell} ||,
\end{align}
where $|| \hat{O} ||$ denotes the spectral norm of the operator $\hat{O}$. Using this, along with the sub-multiplicity of the spectral norm, we have,  
\begin{align}
\label{general_error_bound}
|| \hat{\mathcal{E}}(t,t_0,\tau_M) || \leq 4 \sum_{\alpha,\nu,\ell} ||\hat{X}_{\alpha \ell} ||~|| \hat{X}_{\nu \ell} || \int_{\tau_M}^{\infty}dt   |Q_{\alpha \nu}^{(\ell)}(t)|, 
\end{align}
where, we have additionally extended the upper limit of the integration to infinity.

\subsection{For our system-bath coupling}
Let us now go to a system-bath coupling of the form,
\begin{align}
\hat{\mathcal{H}}_{SB}^{(\ell)}=\sum_{r=1}^\infty(\kappa_{r\ell}\hat{S}^\dagger_\ell \hat{B}_{r\ell} + \kappa_{r\ell}^*\hat{B}_{r\ell}^\dagger\hat{S}_\ell ).
\end{align}This is a slightly more specific system-bath coupling instead of the absolutely general one in Eq.(\ref{general_set_up}). This can be cast in the form of Eq.(\ref{general_set_up}), with the following definitions
\begin{align}
& \hat{X}_{1\ell} = \hat{S}_\ell^\dagger + \hat{S}_\ell,~ \hat{X}_{2\ell} = i(\hat{S}_\ell^\dagger - \hat{S}_\ell) \nonumber \\
& \hat{\mathcal{B}}_{1 \ell} = \sum_{r=1}^\infty \kappa_{r\ell} \frac{\hat{B}_{r\ell}^\dagger + \hat{B}_{r\ell} }{2},~~\hat{\mathcal{B}}_{2 \ell} = i\sum_{r=1}^\infty \kappa_{r\ell} \frac{\hat{B}_{r\ell}^\dagger - \hat{B}_{r\ell} }{2}.
\end{align}
With these definitions, $Q_{\alpha \nu}^{(\ell)}(t)$ are elements of a $2\times 2$ matrix. The elements are given by
\begin{align}
& Q_{11}^{(\ell)}(t) = Q_{22}^{(\ell)}(t)\nonumber \\
&= \frac{1}{4} \int \frac{d\omega}{2\pi} \mathfrak{J}_\ell(\omega) \Big[\big(1\mp \mathfrak{n}_\ell(\omega)\big) e^{-i\omega t} + \mathfrak{n}_\ell(\omega) e^{i\omega t} \Big], \\
&  Q_{12}^{(\ell)}(t) = Q_{21}^{(\ell)*}(t)\nonumber \\
&= \frac{i}{4} \int \frac{d\omega}{2\pi} \mathfrak{J}_\ell(\omega)\Big[\big(1\mp \mathfrak{n}_\ell(\omega)\big) e^{-i\omega t} - \mathfrak{n}_\ell(\omega) e^{i\omega t} \Big], \nonumber
\end{align}
where 
\begin{align}
\label{bath_spectral_functions}
\mathfrak{J}_\ell(\omega) = 2\pi\sum_{r=1}^\infty |\kappa_{r\ell}|^2 \delta (\omega - \Omega_{r\ell})
\end{align} 
is the spectral function of the $\ell$th bath, and $\mathfrak{n}_\ell(\omega)=[{\rm exp}(\beta_\ell(\omega-\mu_\ell)~\pm~1]^{-1}$ is the Fermi or Bose distribution corresponding to the initial state of the $\ell$th bath. Thus, it is clear that, the memory time $\tau_M$ is governed by the time of decay for the Fourier transforms of $\mathfrak{J}_\ell(\omega)$ and $\mathfrak{J}_\ell(\omega)\mathfrak{n}_\ell(\omega)$. For rigorous bounds,  using the sub-additivity of absolute value, we have
\begin{align}
& |Q_{11}^{(\ell)}(t)| = |Q_{22}^{(\ell)}(t)| \leq \frac{1}{4} \Big(~|a_\ell(t)|+2 |b_\ell(t)|~\Big), \nonumber \\
& |Q_{12}^{(\ell)}(t)| = |Q_{21}^{(\ell)}(t)| \leq \frac{1}{4} \Big(~|a_\ell(t)|+2 |b_\ell(t)|~\Big),
\end{align}
where 
\begin{align}
a_\ell(t)=\int \frac{d\omega}{2\pi}\mathfrak{J}_\ell(\omega) e^{i\omega t},~b_\ell(t)=\int \frac{d\omega}{2\pi}\mathfrak{J}_\ell(\omega) \mathfrak{n}_\ell(\omega)e^{i\omega t}.
\end{align}
This simplifies the general error bound in Eq.(\ref{general_error_bound}) to
\begin{align}
\label{simplified_error_bound}
& || \hat{\mathcal{E}}(t,t_0,\tau_M) || \leq  \sum_\ell A_\ell \int_{\tau_M}^\infty dt~\Big(~|a_\ell(t)|+2 |b_\ell(t)|~\Big),
\end{align}
with $A_\ell = \sum_{\alpha, \nu=1}^2  ||\hat{X}_{\alpha \ell} ||~|| \hat{X}_{\nu \ell} ||$, which is a finite positive number. While this bound is general, it is very conservative. In practice, a simpler and better approach is to directly calculate $a_{\ell}(t)$ and $b_{\ell}(t)$, and   choose $\tau_M$ as the time when their magnitude decay below some small percentage of their original value.

\section{NEGF results for NESS of non-interacting systems}\label{app:NEGF}

In the main text, we have used non-equilibrium Green's functions (NEGF) to obtain the exact non-equilibrium steady state (NESS) results for the non-interacting set-up ($V=0$ in Eq.(\ref{example_set_up})). Here we give the relevant formulas for the same.

Any non-interacting system Hamiltonian can be written in the form
\begin{align}
\hat{\mathcal{H}}_S = \sum_{\ell,m=1}^{L_S} \mathbf{H}_{\ell,m} \hat{c}_\ell^\dagger \hat{c}_m,
\end{align}
where $\mathbf{H}$ is a $L_S \times L_S$ Hermitian matrix. For our set-up with $V=0$ in Eq.(\ref{example_set_up})), $\mathbf{H}$ is a tridiagonal matrix where off-diagonal elements are $1$, and diagonal elements are zero, 
\begin{align}
\mathbf{H}_{\ell m} = \delta_{\ell,m-1} + \delta_{\ell-1,m}.
\end{align}
The NEGF is given by
\begin{align}
\mathbf{G}(\omega) = [\omega \mathbb{I} - \mathbf{H}-\Sigma^{(1)}(\omega) - \Sigma^{(2)}(\omega)]^{-1},
\end{align}
where, $\mathbb{I}$ is the $L_S$ dimensional identity matrix, $\Sigma^{(1)}(\omega)$, ($\Sigma^{(2)}(\omega)$) is the self energy matrix of the bath attached to first (last) site. The only non-zero elements of the self-energy matrices are
\begin{align}
& \Sigma^{(1)}_{11}(\omega) = \frac{1}{2} \left(i\mathfrak{J}_1(\omega) + \mathfrak{J}_1^H(\omega)\right), \nonumber \\
& \Sigma^{(2)}_{L_S L_S}(\omega) = \frac{1}{2} \left(i\mathfrak{J}_2(\omega) + \mathfrak{J}_2^H(\omega)\right),
\end{align} 
where $\mathfrak{J}_\ell^H(\omega)$ is the Hilbert transform of $\mathfrak{J}_\ell(\omega)$.

Since the NESS for a non-interacting system is Gaussian, the entire state can be obtained from the correlations of the form $\langle \hat{c}_p^\dagger \hat{c}_q \rangle_{\rm NESS}= {\rm Tr}\left( \hat{c}_p^\dagger \hat{c}_q \rho_{\rm NESS}\right)$, where  $\rho_{\rm NESS}$ is NESS density matrix. These correlations are given in terms of NEGF by
\begin{align}
\langle \hat{c}_p^\dagger \hat{c}_q \rangle_{\rm NESS}=& \int \frac{d\omega}{2\pi} \Big[ \mathbf{G}_{p 1}^*(\omega) \mathbf{G}_{q 1}(\omega) \mathfrak{J}_1(\omega) \mathbf{n}_1(\omega) \nonumber \\
&+ \mathbf{G}_{p L_S}^*(\omega) \mathbf{G}_{q L_S}(\omega) \mathfrak{J}_{L_S}(\omega) \mathbf{n}_{L_S}(\omega) \Big].
\end{align}
The above equation was used to numerically exactly calculate NEGF results.

\section{TEBD in mixed basis}\label{app:TEBD_mixed_basis}

The tensor network technique we have used for time evolution of the open system with finite sized baths is time-evolution-by-block-decimation (TEBD) in mixed basis. Here we give the details of the this technique.

\subsection{Preparing the set-up in the mixed basis}
\begin{figure}
\includegraphics[width=\columnwidth]{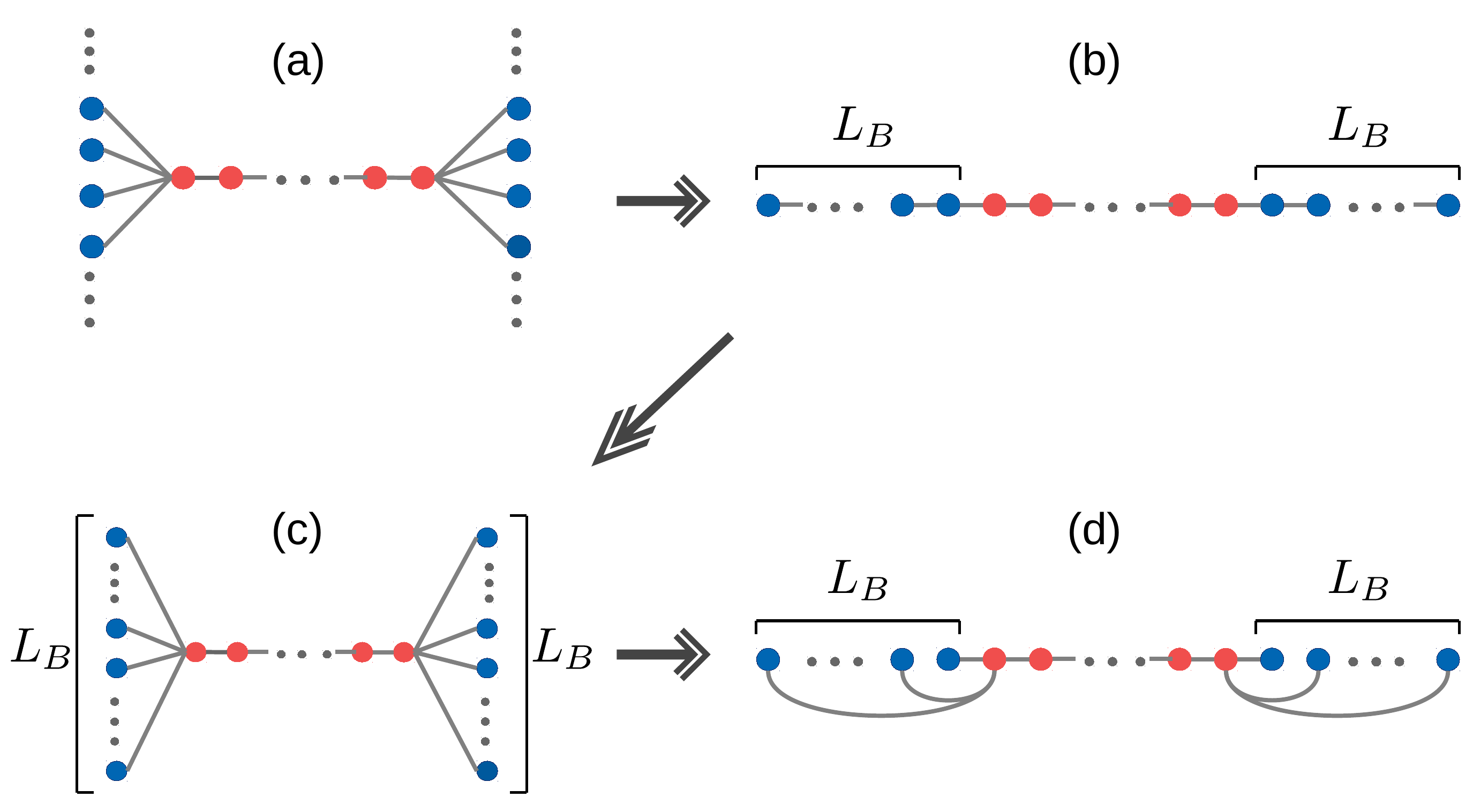} 
\caption{Preparation of the set-up in mixed basis: (a) The initial set-up with baths (blue) consisting of infinite number of modes attached at first and last sites of a system (red); (b) The conversion (rc-mapping) of the baths to one dimensional chains with the first site of the chain attached to the system. The size of the chain is finite, $L_B$ which is chosen to be proportional to the maximum simulation time $\tau$. (c) We move to the single-particle eigenbasis of the baths represented by finite sized chains. (d) We view the set-up as a one-dimensional system with long-range hoppings between the bath modes and the system sites attached to the baths.  }
\label{fig:mixed_basis_conversion} 
\end{figure}

The set-up we have considered for the numerical example is the defined by the following fermionic Hamiltonian $\hat{\mathcal{H}}=\hat{\mathcal{H}}_S+\hat{\mathcal{H}}_{SB}^{(1)}+\hat{\mathcal{H}}_{SB}^{(2)}+\hat{\mathcal{H}}_B^{(1)}+\hat{\mathcal{H}}_B^{(2)}$,
\begin{align}
\label{example_set_up}
 \hat{\mathcal{H}}_S &= \sum_{m=1}^{L_S-1}\left( \hat{c}_m^\dagger \hat{c}_{m+1}+\hat{c}_{m+1}^\dagger \hat{c}_{m} + V \hat{n}_m \hat{n}_{m+1}\right)\nonumber \\
 &+h\sum_{m~{\rm odd}}\hat{n}_m, \nonumber \\
\hat{\mathcal{H}}_{SB}^{(1)}&=\sum_{r=1}^\infty\kappa_{r1}(\hat{c}_1^\dagger \hat{B}_{r1} + \hat{B}_{r1}^\dagger\hat{c}_1 ), \\
\hat{\mathcal{H}}_{SB}^{(2)}&=\sum_{r=1}^\infty\kappa_{r2}(\hat{c}_{L_S}^\dagger \hat{B}_{r2} + \hat{B}_{r2}^\dagger\hat{c}_{L_S} ), \nonumber \\
\hat{\mathcal{H}}_B^{(1)}& =    \sum_{r=1}^{\infty} \Omega_{r1} \hat{B}_{r1}^\dagger \hat{B}_{r1},~\hat{\mathcal{H}}_B^{(2)} =    \sum_{r=1}^{\infty} \Omega_{r2} \hat{B}_{r2}^\dagger \hat{B}_{r2}. \nonumber
\end{align}
In the above Hamiltonian, the baths are modelled by an infinite number of modes. The bath spectral functions are as given in Eq.(\ref{bath_spectral_functions}).
The geometry of the set-up can be pictorially represented as in Fig.~\ref{fig:mixed_basis_conversion}(a).

We recursively use reaction-coordinate (rc) mapping to convert the baths into one dimensional chains with the first site of the chains attached to the system. Further, assuming that the time evolution is up to a time $\tau$, we choose a finite size of the baths, $L_B$, proportional to $\tau$. After recursively using rc mapping, we have,
\begin{align}
& \hat{\mathcal{H}}_B^{(1)} = \sum_{p=1}^{L_B} \left(\varepsilon_{p,1} \hat{b}_{p,1}^\dagger\hat{b}_{p,1} + g_{p,1}(\hat{b}_{p,1}^\dagger\hat{b}_{p+1,1}+\hat{b}_{p+1,1}^\dagger\hat{b}_{p,1})\right),\nonumber \\
& \hat{\mathcal{H}}_B^{(2)} = \sum_{p=1}^{L_B} \left(\varepsilon_{p,2} \hat{b}_{p,2}^\dagger\hat{b}_{p,2} + g_{p,2}(\hat{b}_{p,2}^\dagger\hat{b}_{p+1,2}+\hat{b}_{p+1,2}^\dagger\hat{b}_{p,2})\right), \nonumber \\
& \hat{\mathcal{H}}_{SB}^{(1)}=\gamma_{1}(\hat{b}_{1,1}^\dagger \hat{c}_1  + \hat{c}^\dagger_1 \hat{b}_{1,1}) \\
& \hat{\mathcal{H}}_{SB}^{(2)}=\gamma_{2}(\hat{b}_{1,2}^\dagger \hat{c}_{L_S}  + \hat{c}^\dagger_{L_S} \hat{b}_{1,2}) \nonumber.
\end{align}
The parameters $\gamma_{\ell}$, $\ell=\{1,2\}$  are given by
\begin{align}
\gamma_{\ell}^2 = \frac{1}{2\pi} \int d\omega~\mathfrak{J}_\ell(\omega).
\end{align} 
The on-site potentials $\varepsilon_{p,\ell}$ and the hoppings $g_{p,\ell}$ are obtained from the following set of recursion relations
\begin{align}
& \mathfrak{J}_{p, \ell}(\omega)= \frac{4g_{p-1,\ell}^2 \mathfrak{J}_{p-1,\ell}(\omega) }{\left[\mathfrak{J}_{p-1,\ell}^H (\omega)\right]^2 +\left[\mathfrak{J}_{p-1,\ell}(\omega)\right]^2},  \nonumber \\
& g_{p, \ell}^2 = \frac{1}{2\pi} \int d\omega \mathfrak{J}_{p,\ell}(\omega), \\
& \varepsilon_{p, \ell} = \frac{1}{2\pi g_{p, \ell}^2} \int d\omega~\omega \mathfrak{J}_{p,\ell}(\omega) \nonumber,
\end{align}
with $\mathfrak{J}_{0,\ell}(\omega) =\mathfrak{J}_{\ell}(\omega) $, the index $p$ going from $1$ to $L_B$ and $\mathfrak{J}_{p,\ell}^H (\omega)$ being the Hilbert transform of  $\mathfrak{J}_{p,\ell} (\omega)$,
\begin{align}
\mathfrak{J}_{p,\ell}^H(\omega)= \frac{1}{\pi}\mathcal{P}\int_{-\infty}^{\infty} d\omega^\prime \frac{\mathfrak{J}_{p,\ell}(\omega^\prime)}{\omega -\omega^\prime},
\end{align} 
where $\mathcal{P}$ denotes the principal value \cite{rc_mapping_QTD_book}. With above, we have mapped the infinite bath into the finite-sized chain required for our purpose. The geometry of the set-up is now as in Fig.~\ref{fig:mixed_basis_conversion}(b).

Next we go to the single-particle eigenbasis of the finite-sized chains. For this, we rewrite the bath Hamiltonian as 
\begin{align}
\hat{\mathcal{H}}_B^{(\ell)} = \sum_{p,q=1}^{L_B} \mathbf{H}_{pq}^{(\ell)} \hat{b}_{p,\ell}^\dagger \hat{b}_{q,\ell},
\end{align}
where $\mathbf{H}^{(\ell)}$ is a symmetric tridiagonal matrix with diagonal elements $\{\varepsilon_{p,\ell}\}$ and off-diagonal elements $\{g_{p,\ell}\}$. The annihilation operators in the single-particle eigenbasis are given by 
\begin{align}
\hat{a}_{\alpha,\ell} = \sum_{p=1}^{L_B}\Phi_{p \alpha}^{(\ell)} \hat{b}_{p,\ell},
\end{align}
where $\Phi$ is the matrix that diagonalizes $\mathbf{H}^{(\ell)}$,
\begin{align}
\Phi^{(\ell)^T} \mathbf{H}^{(\ell)} \Phi ^{(\ell)} = \mathbf{D}.
\end{align}
Here, $\mathbf{D}={\rm diag}\{\mathcal{E}_{\alpha\ell}\}$ is a diagonal matrix containing the eigenvalues of the matrix $\mathbf{H^{(\ell)}}$, and $\Phi^{(\ell)^T}$ denotes the transpose of $\Phi^{(\ell)}$. In this basis, the system-bath coupling and the bath Hamiltonians are 
\begin{align}
& \hat{\mathcal{H}}_{SB}^{(1)}=\sum_{\alpha=1}^{L_B}\gamma_1 \Phi_{1 \alpha}^{(1)}(\hat{c}_1^\dagger \hat{a}_{\alpha,1} + \hat{a}_{\alpha,1}^\dagger\hat{c}_1 ), \\
& \hat{\mathcal{H}}_{SB}^{(2)}=\sum_{\alpha=1}^{L_B}\gamma_2 \Phi_{1 \alpha}^{(2)}(\hat{c}_{L_S}^\dagger \hat{a}_{\alpha,2} + \hat{a}_{\alpha,2}^\dagger\hat{c}_{L_S} ), \nonumber \\
& \hat{\mathcal{H}}_B^{(1)} =    \sum_{\alpha=1}^{L_B} \mathcal{E}_{\alpha1} \hat{a}_{\alpha,1}^\dagger \hat{a}_{\alpha,1},~\hat{\mathcal{H}}_B^{(2)} =    \sum_{\alpha=1}^{L_B} \mathcal{E}_{\alpha,2} \hat{a}_{\alpha,2}^\dagger \hat{a}_{\alpha,2}. \nonumber
\end{align}
With this, the geometry of the set-up becomes as in Fig.~\ref{fig:mixed_basis_conversion}(c).

For using TEBD, we need a one-dimensional system. So, we arrange the bath modes in a single line, and view the set-up as a one dimensional system. The added complication becomes that now there is long-range hopping between the bath modes and the system sites attached to the bath modes, as shown in Fig.~\ref{fig:mixed_basis_conversion}(d). To address the long-range hoppings, we need to use TEBD in combination with fermionic swap gates, as we discuss in following subsections.

\subsection{Preparing the gates and the initial state}

With the Hamiltonian obtained in the previous subsection, we now want to calculate
\begin{align}
& \hat{\rho}_{\rm tot}(t) = e^{-i\hat{H}t} \hat{\rho}_{tot}(0) e^{i\hat{H}t}, \nonumber \\ 
& \hat{\rho}_{\rm tot}(0) = \left[\prod_{\alpha=1}^{L_B} \hat{\rho}_{\alpha1}^B \right]\hat{\rho}(0) \left[\prod_{\alpha=1}^{L_B} \hat{\rho}_{\alpha2}^B \right] \\
& \hat{\rho}_{\alpha\ell}^B= \frac{e^{-\beta_\ell (\mathcal{E}_{\alpha\ell}-\mu_\ell) \hat{a}_{\alpha,\ell}^\dagger \hat{a}_{\alpha,\ell}}}{{\rm Tr}\left(e^{-\beta_\ell (\mathcal{E}_{\alpha\ell}-\mu_\ell) \hat{a}_{\alpha,\ell}^\dagger \hat{a}_{\alpha,\ell}}\right)},~\ell=\{1,2\}  \nonumber. 
\end{align}
The above equation highlights that the thermal states of the baths are product states in the mixed basis, which is one of the advantages. We move to the superoperator representation where the density matrix is represented as a vector and the operation of any unitary on it is given as follows
\begin{align}
& \hat{\rho}_{\rm tot}(0) \rightarrow | \hat{\rho}_{\rm tot}(0) \rangle \nonumber \\
& U \hat{\rho}_{\rm tot}(0) U^\dagger \rightarrow U \otimes U^\dagger | \hat{\rho}_{\rm tot} (0) \rangle,
\end{align}
where $\otimes$ denotes Kronecker product. In order to use TEBD, we represent vector corresponding to the initial density matrix as a matrix-product-state (MPS). We take the initial MPS to be completely left-canonicalized. This only requires that the system state is left-canonicalized, because the baths are initially in product state and the system is in product state with the baths.

The next step is to decompose the Hamiltonian into two-site terms. The system Hamiltonian is naturally decomposed as $\hat{\mathcal{H}}_S = \sum_{m=1}^{N_S-1} \hat{\mathcal{H}}_{S_m}$ with $\hat{\mathcal{H}}_{S_m}$ given by
\begin{align}
\hat{\mathcal{H}}_{S_m} = \hat{c}_m^\dagger \hat{c}_{m+1}+\hat{c}_{m+1}^\dagger \hat{c}_{m} + V \hat{n}_m \hat{n}_{m+1}.
\end{align}
The bath and the system-bath coupling Hamiltonians are naturally decomposed into $\hat{\mathcal{H}}^{(\ell)}_{B}=\sum_{\alpha=1}^{L_B} \hat{\mathcal{H}}^{(\ell)}_{B_\alpha}$,
\begin{align}
&\hat{\mathcal{H}}^{(1)}_{B_\alpha}=\mathcal{E}_{\alpha1} \hat{a}_{\alpha,1}^\dagger \hat{a}_{\alpha,1}+\gamma_1 \Phi_{1 \alpha}^{(1)}(\hat{c}_1^\dagger \hat{a}_{\alpha,1} + \hat{a}_{\alpha,1}^\dagger\hat{c}_1 ) \nonumber \\
&\hat{\mathcal{H}}^{(2)}_{B_\alpha}=\mathcal{E}_{\alpha2} \hat{a}_{\alpha,2}^\dagger \hat{a}_{\alpha,2} + \gamma_2 \Phi_{1 \alpha}^{(2)}(\hat{c}_{L_S}^\dagger \hat{a}_{\alpha,2} + \hat{a}_{\alpha,2}^\dagger\hat{c}_{L_S} ).
\end{align}
We Jordan-Wigner transform the above two-site fermionic Hamiltonians,
\begin{align}
& \hat{\mathcal{H}}_{S_m}=\hat{\sigma}_m^{+} \hat{\sigma}^-_{m+1}+\hat{\sigma}_m^{-} \hat{\sigma}^+_{m+1} + V \left(\frac{\mathbf{I}+\hat{\sigma}_m^z}{2}\right)\left(\frac{\mathbf{I}+\hat{\sigma}_{m+1}^z}{2}\right), \nonumber \\
&\hat{\mathcal{H}}^{(1)}_{B_\alpha}=\mathcal{E}_{\alpha1} \left(\frac{\mathbf{I}+\hat{\tau}_{\alpha1}^z}{2}\right)\left(\frac{\mathbf{I}+\hat{\tau}_{\alpha1}^z}{2}\right)
\nonumber\\
&\hspace*{25pt}+\gamma_1 \Phi_{1 \alpha}^{(1)}(\hat{\sigma}_1^+ \hat{\tau}_{\alpha1}^- + \hat{\sigma}_1^- \hat{\tau}_{\alpha1}^+ ), \nonumber \\
&\hat{\mathcal{H}}^{(2)}_{B_\alpha}=\mathcal{E}_{\alpha2} \left(\frac{\mathbf{I}+\hat{\tau}_{\alpha2}^z}{2}\right)\left(\frac{\mathbf{I}+\hat{\tau}_{\alpha2}^z}{2}\right)\nonumber \\
&\hspace*{25pt}+\gamma_1 \Phi_{1 \alpha}^{(2)}(\hat{\sigma}_{L_S}^+ \hat{\tau}_{\alpha2}^- + \hat{\sigma}_{L_S}^- \hat{\tau}_{\alpha2}^+ ),
\end{align}
where $\hat{\sigma}_m^{+,-,z}$ are the usual spin half operators at site $m$ of the system, while $\hat{\tau}_{\alpha \ell}^{+,-,z}$ are the corresponding ones for the $\alpha$th bath mode of the $\ell$th bath and $\mathbb{I}$ is the identity operator at the corresponding site. It is important to note that in $\hat{\mathcal{H}}^{(1)}_{B_\alpha}$ and $\hat{\mathcal{H}}^{(2)}_{B_\alpha}$ the operator representing system site is written to the left of that of the bath site. This convention is to be maintained in the following. The following superoperator gates for time evolution by a  Trotterized time step $dt/2$ are constructed from the Jordan-Wigner transformed two-site Hamiltonians
\begin{align}
& U_m = e^{-i\hat{\mathcal{H}}_{S_m}{dt}/{2}} \otimes e^{i\hat{\mathcal{H}}_{S_m}{dt}/{2}} \nonumber \\
& U^{1B}_\alpha = e^{-i\hat{\mathcal{H}}_{B_\alpha}^{(1)}{dt}/{2}} \otimes e^{i\hat{\mathcal{H}}_{B_\alpha}^{(1)}{dt}/{2}}  \\
& U^{2 B}_\alpha = e^{-i\hat{\mathcal{H}}_{B_\alpha}^{(2)}{dt}/{2}} \otimes e^{i\hat{\mathcal{H}}_{B_\alpha}^{(2)}{dt}/{2}} \nonumber. 
\end{align}
To account for fermionic anti-commutation relations, we need to use two-site fermionic swap gates, given by the following matrix in the computational basis
\begin{align}
S = \left( 
\begin{array}{cccc}
1 & 0 & 0 & 0 \\
0 & 0 & 1 & 0 \\
0 & 1 & 0 & 0 \\
0 & 0 & 0 & -1
\end{array}
\right).
\end{align}
\begin{figure}
\includegraphics[width=\columnwidth]{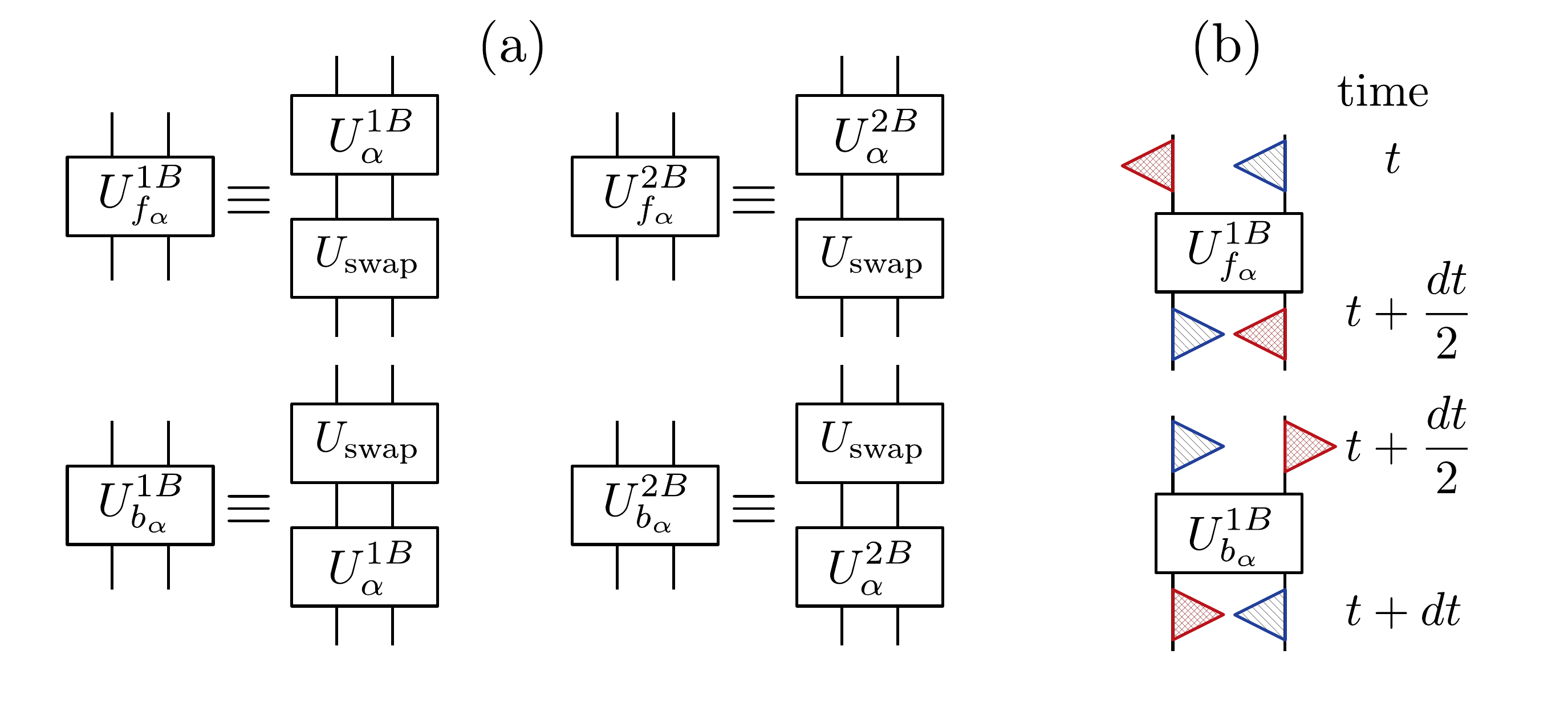} 
\caption{(a) The figure gives pictorial representation of the composite gates for system-bath evolution. (b) The figure shows the action of two of the composite gates. The other two composite gates operate similarly. Here the red triangle represents MPS for the system site, and the other triangle represents MPS for the bath site. The triangle pointing left represents right-canonicalized tensor and the triangle pointing right represents left-canonicalized tensor. It is important to note that  $U^{1B}_{f_\alpha}$ operates when the system site is to the left of the bath site, and $U^{1B}_{b_\alpha}$ operates when the bath site is to the left of the system site.}
\label{fig:composite_gates} 
\end{figure}
\begin{figure*}
\includegraphics[width=\textwidth]{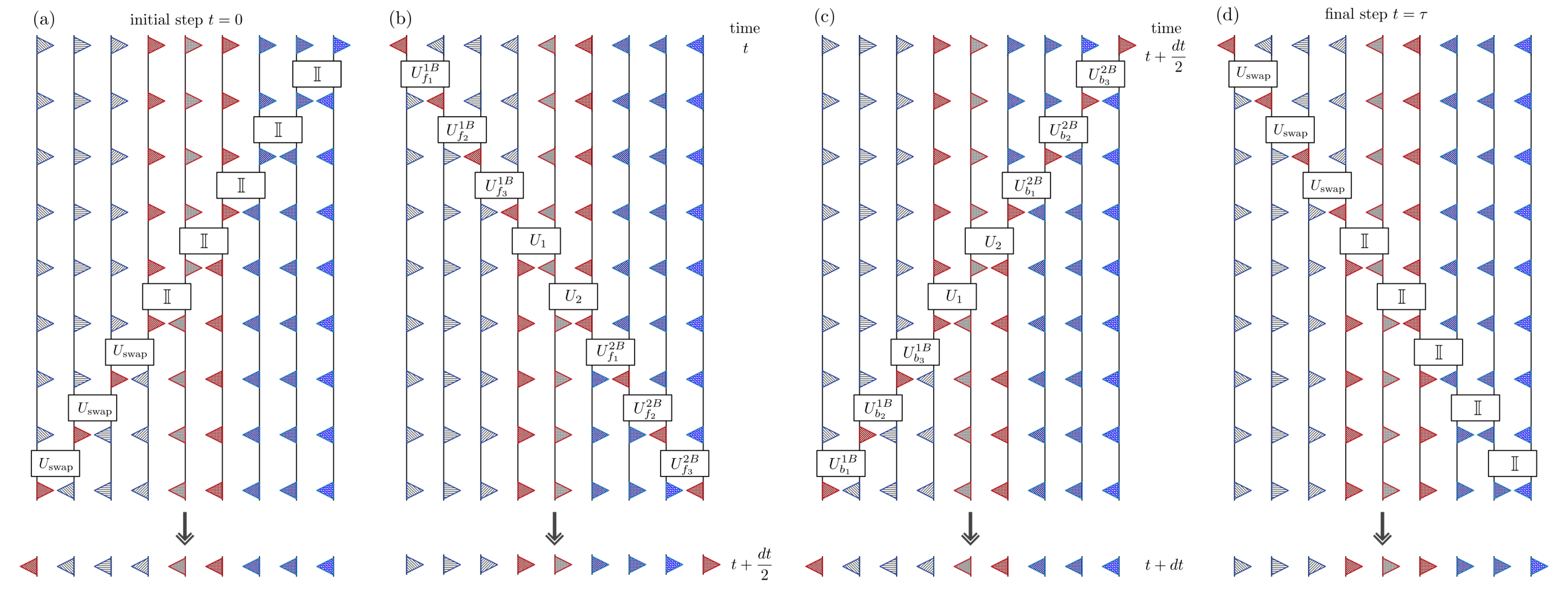} 
\caption{The steps for TEBD are shown for the case of three sites in system (red triangles) and three sites in each bath. The triangles pointing left represent right-canonicalized tensors, the triangles pointing right represent left-canonicalized tensors. The last step of each panel involves left or right canonicalization of the first or last tensors, as the case may be. Panel (a) gives the initial step which takes a fully left-canonicalized MPS and  shifts the first system site to the left end of the chain while making the MPS fully right-canonicalized. Panels (b) and (c) together give the time evolution by a time-step $dt$. The MPS at the beginning and at the end of the time evolution is a fully right-canonicalized state with the first system site shifted to the left end of the chain. Panel (d) gives the final step of which takes a fully right-canonicalized state with the first system site shifted to the left end of the chain and converts it into a completely left canonicalized with the original placement of system sites.  }
\label{fig:TEBD_mixed_basis} 
\end{figure*}
The superoperator representing the swap gate is obtained by
\begin{align}
U_{swap} = S \otimes S.
\end{align}
To efficiently carry out the TEBD algorithm in presence of the long-ranged hoppings we will need to define the following  four kinds of composite gates
\begin{align}
& U^{1B}_{f_\alpha} = \left(S  e^{-i\hat{\mathcal{H}}_{B_\alpha}^{(1)}{dt}/{2}}\right) \otimes \left(e^{i\hat{\mathcal{H}}_{B_\alpha}^{(1)}{dt}/{2}} S\right), \nonumber \\
& U^{2B}_{f_\alpha} = \left(S  e^{-i\hat{\mathcal{H}}_{B_\alpha}^{(2)}{dt}/{2}}\right) \otimes \left(e^{i\hat{\mathcal{H}}_{B_\alpha}^{(2)}{dt}/{2}} S \right), \\
& U^{1B}_{b_\alpha} =   \left(e^{-i\hat{\mathcal{H}}_{B_\alpha}^{(1)}{dt}/{2}} S \right) \otimes \left( S e^{i\hat{\mathcal{H}}_{B_\alpha}^{(1)}{dt}/{2}} \right),  \nonumber \\
& U^{2B}_{b_\alpha} =   \left(e^{-i\hat{\mathcal{H}}_{B_\alpha}^{(2)}{dt}/{2}} S \right) \otimes\left(  S e^{i\hat{\mathcal{H}}_{B_\alpha}^{(2)}{dt}/{2}} \right).  \nonumber 
\end{align}
The pictorial representation of these composite gates as well as their action on a two-site MPS consisting of a system and a bath site, are shown in Fig.~\ref{fig:composite_gates}. As will see below, after a special ordering of the sites, we use the above composite gates for the time evolution via TEBD.  This makes the cost of simulating this set-up with long-ranged hoppings the same as that of one with only nearest neighbour hoppings.

\subsection{The time evolution}
We will describe the time evolution keeping in mind a hypothetical example, where the system has three sites, and each bath also has three sites (Fig.~\ref{fig:TEBD_mixed_basis}). The time evolution consists of three steps.
\paragraph*{The initial step---}
The initial density matrix is taken in fully left-canonicalized form.  The goal of the initial step is to shift the first site of the system to the left end of the chain while converting the MPS into fully right-canonicalized form. This is to be done to do away with essentially all overhead costs of having long-ranged system-bath hoppings, as we will see below. This step is achieved by operating with identity gates sequentially starting from the right end of the chain up to the first system site, and then operating on the rest of the sites by swap gates (see Fig.~\ref{fig:TEBD_mixed_basis}(a)). Finally the first tensor of the MPS is right-canonicalized.
\paragraph*{The time evolving step---}
The time evolution by a Trotterized time step of $dt$ is done in two time steps of $dt/2$. The first step starts with a fully right-canonicalized MPS with the first system site at the left end of the chain, and operates the following composite system-bath gates and the system gates sequentially from the left end to the right end (see Fig.~\ref{fig:TEBD_mixed_basis}(b)),
\begin{align}
\left(\prod_{\alpha=1}^{L_B} U^{2B}_{f_\alpha}\right)\left(\prod_{m=1}^{{L_S}-1}U_m \right) \left(\prod_{\alpha=1}^{L_B} U^{1B}_{f_\alpha}\right) | \hat{\rho}_{\rm tot}(t) \rangle.
\end{align}
At the end of this, the last tensor of the MPS is left canonicalized. The resulting MPS is a fully left-canonicalized one with the last site of the system shifted to the right end of the chain, and the first site of the system restored to its original position. In the next step, the following gates are operated sequentially from the right end to the left end on this MPS (see Fig.~\ref{fig:TEBD_mixed_basis}(c)),
\begin{align}
\left(\prod_{\alpha=L_B}^{1} U^{1B}_{b_\alpha}\right) \left(\prod_{m={L_S}-1}^{1}U_m \right) \left(\prod_{\alpha=L_B}^{1} U^{2B}_{b_\alpha}\right) | \hat{\rho}_{\rm tot}(t+\frac{dt}{2}) \rangle.
\end{align}
At the end of this the first tensor is right-canonicalized. The resulting MPS is of the same form as was before the above two operations, but now representing the density matrix after a time evolution by $dt$. For time evolution up to time $t=\tau$, the above two time evolution steps are repeated $\tau/dt$ times.
\paragraph*{The final step---}
The final step at time $t=\tau$ takes a fully right-canonicalized MPS with the first system site shifted to the left end of the chain and restores the first system site to its original position while converting the MPS to fully left-canonicalized representation. This is achieved by operating swap gates sequentially starting from the left $L_B$ times and then operating identity gates sequentially on the rest of the sites (see Fig.~\ref{fig:TEBD_mixed_basis}(d)). Finally, the last tensor of the chain is left-canonicalized. 

\begin{figure}
\includegraphics[width=0.7\columnwidth]{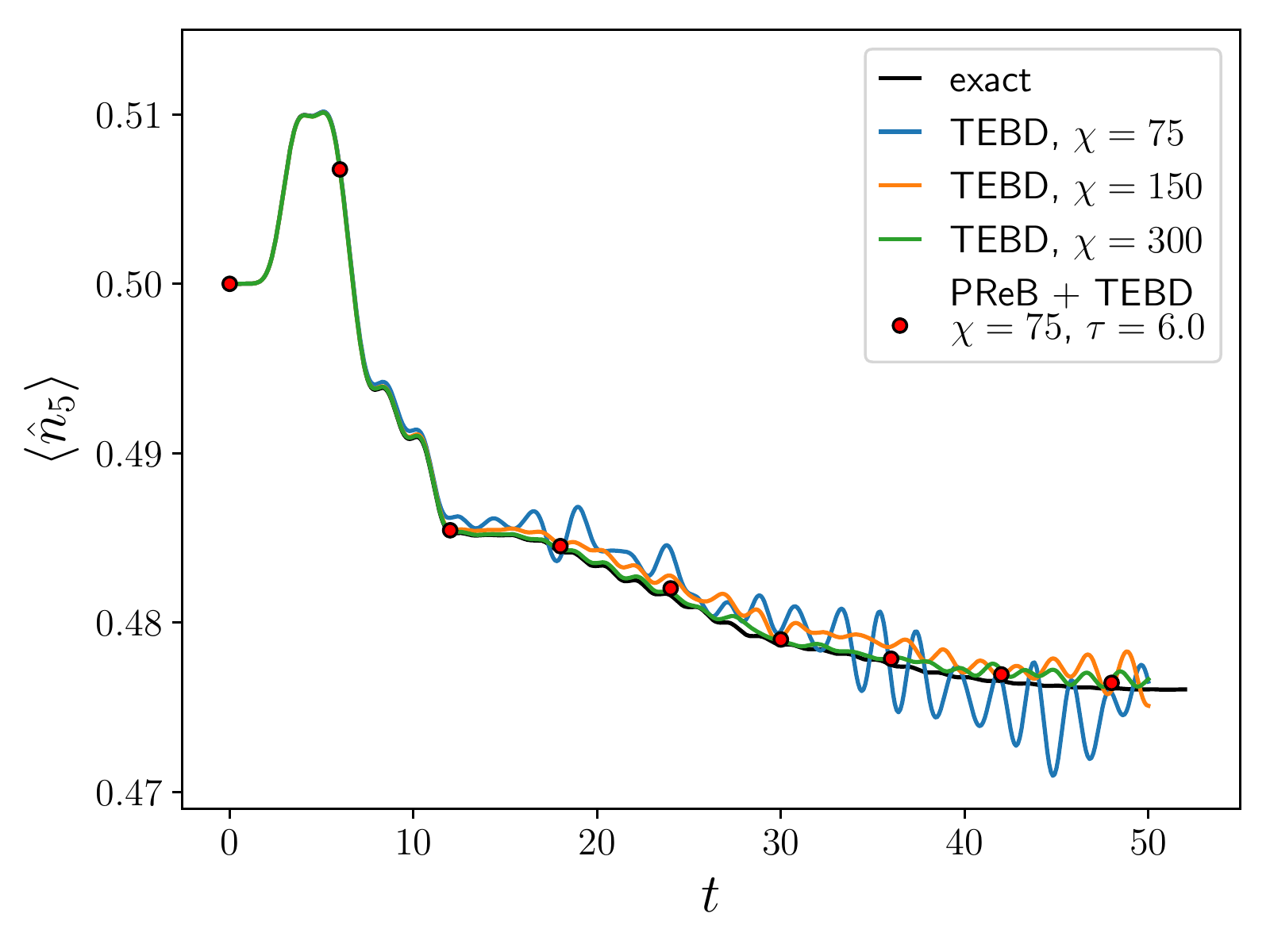} 
\caption{The result from TEBD implementation of the non-interacting open system is compared with exact results without TEBD, and with a PReB+TEBD approach with $\tau=6$ (correspondingly, $L_B=14$). Parameters: $\beta_1=0.1$, $\beta_2=0.2$, $\mu_1=1.5$, $\mu_2=-1.5$, $g_B=2$, $\Gamma_1=1$, $\Gamma_2=2$. The Trotter time step for TEBD is $0.1$. All energy scales are in units of system hopping parameter.  }
\label{fig:non_int_TEBD} 
\end{figure}

The time evolving step above is exactly similar to what it would have been had the chain had only nearest neighbour connections \cite{tensor_network_review_2019}. Thus, by rearranging the sites in the initial step and by using the composite gates for system-bath evolution, we have been able to completely do away with any additional overhead of having long-range system-bath hoppings. Further, the final step is mainly required for PReB calculations, and can be avoided in the continuous time evolution. For PReB calculations, after the final step, the MPS for the density matrix of the system can be obtained by tracing out the baths using standard tensor network techniques \cite{Scholl05}. The MPS of the system density matrix so obtained will be in left-canonicalized form and can be directly used as the initial state for the next iteration of the PReB calculation.

In Fig.~\ref{fig:non_int_TEBD} we compare our mixed-basis TEBD results for various bond dimensions and the PReB+TEBD result with $\tau=6$, with the exact results for the the non-interacting system up to time $t=50$. In the TEBD implementation, to converge up to time $t=50$, we require a bond dimension $\chi=300$. In contrast, the PReB+TEBD approach requires a bond dimension of only $\chi=75$ for convergence.

\bibliography{ref_PReB2}
\end{document}